\documentclass[aps,prd,twocolumn,nofootinbib,superscriptaddress]{revtex4-1}
\usepackage[utf8]{inputenc}

\usepackage{color}
\usepackage{graphicx}
\usepackage{amsmath}
\usepackage{amssymb}
\usepackage{bm}
\usepackage{acronym}
\usepackage{ifthen}
\usepackage{blindtext}
\usepackage[normalem]{ulem}
\usepackage{hyperref}
\usepackage[capitalise]{cleveref}
\usepackage{etoolbox}
\usepackage{xspace}
\usepackage{textcomp}
\usepackage{multirow}
\usepackage{lineno}
\usepackage{tabularx}
\usepackage{subfigure}
\usepackage{orcidlink}

\def\gwh{GW\xspace}

\def\gws{GWs\xspace}

\def\nss{neutron stars\xspace}

\def\bh{black hole\xspace}

\def\dm{DM\xspace}
\def\dmh{DM\xspace}
\def\pbh{PBH\xspace}
\def\pbhs{PBHs\xspace}

\def\bbh{binary black hole\xspace}

\def\fh{\emph{frequency-Hough}\xspace}
\def\lvk{LIGO, Virgo and KAGRA\xspace}
\def\ifo{interferometer\xspace}
\def\ifos{interferometers\xspace}

\def\invkpccubedyr{kpc$^{-3}$yr$^{-1}$\xspace}
\def\pn{PN\xspace}
\def\mf{matched filtering\xspace}
\def\psd{power spectral density\xspace}
\def\nn{\nonumber}

\def\cbcs{compact binary coalescences\xspace}
\def\cbc{compact binary coalescence\xspace}
\def\sfdbs{short fast Fourier transform databases\xspace}

\def\gfh{\emph{Generalized frequency-Hough}\xspace}

\def\OFourA{O4a\xspace}

\def\pmap{\emph{peakmap}\xspace}
\def\pmaps{\emph{peakmaps}\xspace}
\def\erfc{\mathrm{erfc}}

\newcommand{\bea}{\begin{eqnarray}}
\newcommand{\eea}{\end{eqnarray}}
\newcommand{\be}{\begin{equation}}
\newcommand{\ee}{\end{equation}}
\newcommand{\thetathr}{\theta_\text{thr}}

\newcommand{\avgVT}{\ensuremath{\left\langle VT \right\rangle}}

\newcommand{\TFFT}{T_\text{FFT}}
\newcommand{\fgw}{f_\text{GW}}
\newcommand{\fdotgw}{\dot{f}_{\rm GW}}

\newcommand{\fdot}{\dot{f}}
\newcommand{\Tobs}{T_\text{obs}}
\newcommand{\Tpm}{T_\text{PM}}
\newcommand{\fpbh}{f_\text{PBH}}

\newcommand{\ftilde}{\tilde{f}}
\newcommand{\ecc}{e}
\newcommand{\fsup}{f_\text{sup}}
\newcommand{\fmone}{f(\ln m_1)}
\newcommand{\fmtwo}{f(\ln m_2)}
\newcommand{\Mc}{\mathcal{M}}
\newcommand{\tcoal}{t_\text{coal}}
\newcommand{\fmin}{f_\text{min}}
\newcommand{\fmax}{f_\text{max}}
\newcommand{\ssm}{sub-solar mass\xspace}
\newcommand{\msun}{\ensuremath{M_\odot}\xspace}
\newcommand{\crthr}{CR_\text{thr}}

\newcommand{\mpbh}{M_\text{PBH}}

\newcommand{\dnf}{d_{\text{max,95\%}}}

\newtoggle{fullauthorlist}
\toggletrue{fullauthorlist}
\newtoggle{endauthorlist}
\toggletrue{endauthorlist}

\newcommand{\tcr}{}
\newcommand{\tcb}{}

\newcommand{\mconfigi}{\ensuremath{M^\prime_i}\xspace}
\newcommand{\fconfigsi}{\ensuremath{f^\prime_{0,i}}\xspace}

\graphicspath{{./figures/}}

\begin{document}
\title{Search for planetary-mass ultra-compact binaries using data from the first part of the LIGO--Virgo--KAGRA fourth observing run}
\date{\today}

\iftoggle{endauthorlist}{
  %
  %
  \let\mymaketitle\maketitle
  \let\myauthor\author
  \let\myaffiliation\affiliation
  \author{The LIGO Scientific Collaboration}
  \author{The Virgo Collaboration}
  \author{The KAGRA Collaboration}
  \email{Full author list given at the end of the article.}
\noaffiliation
}{
\iftoggle{fullauthorlist}{
  \input{KAGRA-LSC-Virgo-Authors-Aug-2020-aas}
}
{
  \author{The LIGO Scientific Collaboration}
  \affiliation{LSC}
  \author{The Virgo Collaboration}
  \affiliation{Virgo}
  \author{The KAGRA Collaboration}
  \affiliation{KAGRA}
  
}
}


\begin{abstract}

We present a search for gravitational waves from inspiraling, planetary-mass ultra-compact binaries using data from the first part of the fourth observing run of LIGO, Virgo and KAGRA. Finding no evidence of such systems, we determine the maximum distance reach for such objects and their merger rate densities. 
Then, we identify classes of primordial black-hole mass distributions for which these rate limits can be translated into relevant constraints on the mass distribution of primordial black holes, assuming that they compose all of dark matter, in the mass range $[10^{-6},10^{-3}]M_\odot$.  
\tcr{Our constraints for the class of primordial-black-hole mass distribution functions $f(m)$ we consider reach down to $f(m)\sim 0.1$. They are consistent with existing microlensing results in the planetary-mass range, provide a complementary probe to sub-solar mass objects, and are publicly available \cite{LIGO_P2500686}}.
\end{abstract}
        
        \maketitle

\section{Introduction}\label{sec:intro}

{Black holes can form in the Universe from the core collapse of stars, or the merging of neutron stars or black holes.  Since 2015, numerous gravitational-wave (\gwh) observations has allowed us to probe these formation channels and has revealed unexpected features ~\cite{Abbott:2016blz,TheLIGOScientific:2016pea,Abbott:2016nmj,Abbott:2017vtc,Abbott:2017oio,Abbott:2017gyy,LIGOScientific:2018mvr,LIGOScientific:2020stg,Abbott:2020uma,Abbott:2020khf,Abbott:2020tfl,Abbott:2020mjq}, such as low effective spins, black holes in the low-mass gap ([3,5]$\,\msun$) \cite{LIGOScientific:2024elc}, in the pair-instability mass gap \cite{LIGOScientific:2020iuh} or with very unequal mass ratios \cite{LIGOScientific:2020ibl,LIGOScientific:2021djp}. Though a fraction of these could come from primordial black holes (\pbhs), formed through the gravitational collapse of inhomogeneities in the early Universe \cite{Hawking:1971ei,Carr:2019kxo,Bird:2016dcv,Clesse:2016vqa,Sasaki:2016jop}, uncertainties in the astrophysical and primordial formation scenarios and rates are too large to be able to disentangle the two populations.

A relatively unambiguous way to discover a \pbh would be to detect a \ssm \bh.}  {Matched filtering has been used to search for \pbhs between $[0.1,1]\,\msun$ \tcb{in equal-mass systems and down to $0.01\msun$ in asymmetric mass-ratio systems}, which has resulted in upper limits on the abundance of \pbhs in the Universe \cite{LIGOScientific:2019kan,LIGOScientific:2021job,Phukon:2021cus,Nitz:2021vqh,Nitz:2021mzz,LIGOScientific:2022hai,Nitz:2022ltl}. Probing below $0.1\,\msun$, however, has been challenging for \mf because longer signal durations  -- $\mathcal{O}(\rm hours-days)$ versus $\mathcal{O}(100\text{ s})$-- lead to an insurmountable number of templates to analyze. These difficulties have motivated the development of time-frequency domain methods to probe \pbhs with masses between $[10^{-7},10^{-2}]\,\msun$ \cite{Miller:2020kmv,Andres-Carcasona:2023zny,Alestas:2024ubs,Andres-Carcasona:2024jvz,Miller:2024jpo,Miller:2024rca}, which have been used \cite{Miller:2024fpo} to obtain the \gwh constraints on planetary-mass \pbhs using data from the third observing run of \lvk \cite{2015CQGra..32g4001L,2015CQGra..32b4001A,KAGRA:2020tym}. Another possibility is to search for the continuous, almost monochromatic \gwh signal from the early inspiral of planetary-mass \pbh binaries within our galaxy \cite{Miller:2021knj,KAGRA:2022dwb,Miller:2024khl}.}  

In this paper, we use data of the first part of the fourth observing run of \lvk  to search for \gws from inspiraling \pbh binaries.  This work advances beyond previous studies not only through the use of new data, but also by 
identifying {some relevant classes of} \pbh mass functions 
{for which we obtain a significant constraint on the fraction of dark matter (\dm) these \pbhs could compose.}
{Additionally, we consider the impact of binary eccentricity 
on our constraints.

\section{The signal} \label{sec:sig}

Inspiraling compact objects will lose orbital energy via \gwh emission, causing their orbit to shrink over time. When the objects are far from merger, we can write how quickly the \gwh frequency changes over time, i.e. the spin-up $\fdot$, as \cite{maggiore2008gravitational}:

\begin{align}
    \fdotgw&=\frac{96}{5}\pi^{8/3}\left(\frac{G\Mc}{c^3}\right)^{5/3} \fgw^{11/3} \nonumber \\
    &\simeq 1.25\times 10^{-4}\text{ Hz/s} \left(\frac{\Mc}{10^{-3}\,\msun}\right)^{5/3}\left(\frac{\fgw}{100\text{ Hz}}\right)^{11/3},
    \label{eqn:fdot_chirp}
\end{align}
where $\Mc\equiv\frac{(m_1m_2)^{3/5}}{(m_1+m_2)^{1/5}}$ is the chirp mass of the system composed of objects with component masses $m_1,m_2$, $\fgw$ is the \gwh frequency, $c$ is the speed of light, and $G$ is Newton's gravitational constant.  
We can then integrate \cref{eqn:fdot_chirp} over time to obtain how the \gwh frequency changes with time $t$:

\begin{equation}
\fgw(t)=f_0\left[1-\frac{8}{3}kf_0^{8/3}(t-t_0)\right]^{-\frac{3}{8}}~,
\label{eqn:powlaws}
\end{equation}
where $t_0$ is a reference time for the \gwh frequency $f_0$, and $k\equiv\frac{96}{5}\pi^{8/3}\left(\frac{G\Mc}{c^3}\right)^{5/3}$ is a proportionality constant.
\cref{eqn:fdot_chirp} contains only the leading-order term in the post-Newtonian (\pn) expansion. However, the impact of higher-order \pn corrections on the signal's time–frequency evolution is negligible for the non-spinning, widely separated binaries considered here. We discuss how this approximation affects our upper limits in \cref{sec:ul}.

The  amplitude $h_0(t)$ of the \gwh signal also evolves with time \cite{maggiore2008gravitational}:

\begin{align}
h_0(t)&=\frac{4}{d}\left(\frac{G \Mc}{c^2}\right)^{5/3}\left(\frac{\pi \fgw(t)}{c}\right)^{2/3} \nonumber \\
&\simeq 2.56\times 10^{-23}\left(\frac{1\text{ kpc}}{d}\right)\left(\frac{\Mc}{10^{-3}\,\msun}\right)^{5/3}\left(\frac{\fgw(t)}{100\text{ Hz}}\right)^{2/3},
\label{eqn:h0}
\end{align}
where $d$ is the luminosity distance to the source.

To determine how long the inspiraling systems considered in this paper will emit \gws, we can invert \cref{eqn:powlaws} to obtain the time to coalescence $\tcoal$ and let $\fgw(t)\rightarrow\infty$:

\begin{align}
    \tcoal&\simeq\frac{5}{256}\left(\frac{1}{\pi f_0}\right)^{8/3}\left(\frac{c^3}{G\Mc}\right)^{5/3}\nonumber \\
    &\simeq 3.4\text{ days} \left(\frac{100\text{ Hz}}{f_0}\right)^{8/3} \left(\frac{10^{-3}\,\msun}{\Mc}\right)^{5/3}.
    \label{eqn:tcoal}
\end{align}
We term these intermediate-duration signals lasting hours-days ``transient continuous waves''\footnote{This term was coined originally in the context of signals arising after pulsar glitches \cite{Prix:2011qv}.}, whose chirp masses would range from $\mathcal{O}(10^{-5}-10^{-2})\,\msun$. Systems whose chirp masses are less than $10^{-5}\,\msun$ could spend years in the LIGO--Virgo--KAGRA frequency band, and are termed ``continuous waves'' -- analogous to canonical \gwh emission from nonaxisymmetric rotating \nss \cite{Riles:2022wwz} -- and were also searched for in this dataset \cite{o4a_all_sky}.

\section{The search} \label{sec:search}

\subsection{Data}

We consider data from the first part of the fourth observing run of the LIGO Livingston (L1) and Hanford (H1) detectors, called \OFourA\footnote{We used the channel \textsc{GDS--CALIB\_STRAIN\_CLEAN\_AR} with \textsc{CAT1} vetoes \cite{LIGO:2024kkz}.}. This observing run is more sensitive than previous ones across the full frequency band, but particularly at high frequencies \cite{LIGOO4Detector:2023wmz,membersoftheLIGOScientific:2024elc,Capote:2024rmo}. Data were collected between 24 May 2023 15:00:00 UTC and 16 January 2024 16:00:00 UTC, with L1 and H1 online for 69\% and 67.5\% of that time, respectively. The data are calibrated \cite{Karki:2016pht,Viets:2017yvy,Wade:2025tgt} such that, at worst, amplitude and phase uncertainties at $1\sigma$ are 10\% and 10 degrees, respectively. We used data from times at which the \ifos were in ``science mode'' \cite{GoetzRiles:T2400058:v1}. Virgo did not operate during \OFourA, while KAGRA observed for one month. 

The data structures that we used as the input to this search are called \sfdbs \cite{Astone:2005fj,Astone:2025sfdb}, which contain frequency-domain representations of the data every $\sim 17$ minutes that are cleaned of short time-domain disturbances (``glitches'') \cite{Astone:2005fj}. Because we require the fast Fourier transform length $\TFFT\ll17$ minutes, we inverse Fourier transform the data to the time domain and create new time-frequency representations of the data with the desired $\TFFT$.

\subsection{Method}\label{subsec:meth}

{Standard searches for \cbcs use \mf to coherently match signal waveforms to \gwh data. Such analyses work well for short-duration signals, but for the long-duration signals present here, \mf would be computationally infeasible; thus, we employ a \emph{semi-coherent} approach. We break the data into coherent chunks of length $\TFFT$, and sum the power across different FFTs incoherently. Such \emph{semi-coherent} searches require a choice of $\TFFT$ that ensures that the \gwh frequency is monochromatic during each FFT. Practically, $\TFFT$ is a function of $\Mc$, $\fgw$ and the signal duration $\Tpm$, as described in \cref{subsec:parmspace}. }

Here, we use the \gfh to search for \gws from inspiraling compact objects \cite{Miller:2018rbg,Miller:2020kmv,Miller:2024jpo,Menon:2025wce,Pierini:2025gfhv2,miller_2025_17984228}. For a given power-law model of the time-frequency evolution of the signal (\cref{eqn:powlaws}), the \gfh operates on a time-frequency representation of the data called the \pmap (PM), a collection of ones and zeros that indicates the particular frequencies whose power is a local maximum and greater than a chosen threshold. This method performs a transformation from the $t-\fgw$ plane of each detector separately to the $f_0-\Mc$ plane of the source. In doing this transformation, we sum over different possible tracks in the time-frequency plane to accumulate the ``peaks'' (the ones) corresponding to specific chirp masses and coalescence times. The number of unique tracks to search over depends on our analyses parameters -- $\TFFT,\Tpm$ -- and $\fgw$.

In contrast to previous searches, we use \gfh-v2 that is approximately an order of magnitude faster than previous versions \cite{Menon:2025wce,Pierini:2025gfhv2,miller_2025_17984228}, permitting us to perform an extensive follow-up campaign of significant time-frequency tracks.

\subsection{Parameter space construction}\label{subsec:parmspace}

We perform the search in a number of ``configurations'', where each configuration corresponds to a particular frequency band $[\fmin,\fmax]$, $\TFFT$, a \pmap duration $\Tpm$ and a chirp mass range $[\Mc_{\rm min},\Mc_{\rm max}]$. These parameters are chosen to maximize the sensitivity, i.e. the distance reach \cref{eqn:dmax}, towards systems with chirp masses between $[\Mc_{\rm min},\Mc_{\rm max}]$ by considering the changing noise power spectral density of the \ifos, the varying amplitude of the signal over time, and the steep increase in $\fdot$ over time, which results in a decrease of the analysis coherence time if we wish to confine the signal power to one frequency bin during $\TFFT$. By looping over all possible chirp masses and \gwh frequencies, we can empirically determine these parameters by calculating the distance reach, and grouping nearby points in the parameter space to be searched for in a single \pmap with the criterion that no more than 10\% sensitivity is lost with respect to creating separate \pmaps for each of those nearby points. To limit the computational cost of the search, we require that the chosen parameters would lead to detectable signals at least 0.1 kpc from us, and the Doppler shift induced by the relative motion of the Earth and source is confined to one frequency bin for the signal duration.

At the end of this process, we obtain 685 unique configurations with $\TFFT\in[2,13]$ s and $\Tpm\in[2.3\text{ h},6.75\text{ d}]$ covering systems with $\Mc\in [10^{-5},10^{-2}]\,\msun$.

\subsection{Results}\label{subsec:results}

We perform the \gfh on each \pmap in each configuration per detector separately, across the whole observing run $\Tobs\simeq8$ months. Candidates returned from each detector with similar chirp masses and coalescence times are considered ``significant'' and ``in coincidence'' if (1) their parameters do not differ by more than three bins in the two-dimensional $f_0-\Mc$ parameter space, (2) their detection statistics, called the ``critical ratio'' ($CR$), are comparable (within 20\% of each other), (3) their average $CR$ exceeds a threshold of $\sim 7$ determined by the trials factor and assuming Gaussian noise, and (4) their time-frequency tracks do not overlap by more than 30\% with a known noise line \cite{LIGO:2024kkz,GoetzEtAl:T2400204:v2}. \tcr{The $CR$ threshold is set taking into account computing power constraints: the follow-up can be an expensive portion of the search, because the demodulation procedure, discussed below, is very sensitive to parameter uncertainties and $\TFFT$ \cite{Astone:2014esa,Miller:2018rbg}. }

The chosen criteria lead to a manageable $5\times 10^5$ candidates. These candidates were subject to a follow-up procedure in which the original time-series data were demodulated based on the expected phase evolution of the signal, obtained by integrating \cref{eqn:powlaws}. Such a demodulation will ideally ensure the signal remains monochromatic during its duration, thus permitting us to use a longer coherence time. 

We then compute a new time-frequency \pmap after doubling $\TFFT$, and apply the \fh \cite{Astone:2014esa,Piccinni:2018akm,AstoneEtAl:2025exampleFH}, which searches over residual $\fdotgw$ that may occur for imperfect corrections. Similarly to the \gfh, the \fh maps points in the $t-\fgw$ plane of the detector to lines in the $f_0-\fdotgw$ plane of the source. We \tcr{only} require that the new $CR$ exceeds the previously found $CR$ \tcr{by any amount}, which would occur for a real \gwh signal because the coherence time has been increased.

The parameters returned by the \fh are then used to correct the \pmap for any residual modulations, which would lead to a monochromatic signal; then, the \pmap is projected onto the frequency axis. \tcr{At this stage, a new $CR$ is calculated on the projected \pmap, and that $CR$ must exceed the previously calculated $CR$ in order for a candidate to be subject to further analysis.} Nineteen candidates survived one doubling of $\TFFT$ until the \pmap projection, after which all were vetoed.  {See \cref{app:final-search} for more details regarding the final stages of the search and the threshold selection, \tcr{and previous methods papers for graphical descriptions of the follow-up methods, e.g. \cite{Miller:2018rbg,Miller:2019xos,Miller:2024jpo}}.}

\section{Upper limits}\label{sec:ul}

\subsection{Computing expected distance reach}

The procedure for semi-analytically computing upper limits on the
search’s distance reach is described elsewhere \cite{Miller:2024jpo} and was used in \cite{Miller:2024fpo}, so we highlight only the important aspects here and provide more details in \cref{app:ul}. {Although all coincident candidates have been vetoed, we can still use their $CR$ values as thresholds to estimate the distance from which a real signal could have originated in order to not produce a $CR$ larger than that observed in H1 and L1. {If a \gwh arising from a source a certain distance away can pass these thresholds at least 95\% of the time, we can set an upper limit on the distance reach for sources similar to that particular candidate $\dnf$ at which 95\% of injections would be detected, i.e. at a particular time, frequency range, and chirp mass.} We can compute $\dnf$ in two ways: (1) by requiring that an injected \gwh signal in the data be recovered with a $CR$ larger than that returned in both H1 and L1 95\% of the time, or (2) by analytically calculating the distance reach as a function of the coincident candidates returned in the search. We choose to do the latter, but verify that both approaches produce consistent $\dnf$ upper limits -- see \cref{app:ul}.}

For each of the coincident candidates in each configuration before thresholding the detection statistic, we compute $\dnf$. To be conservative, we use the maximum detection statistic of coincident candidates returned in H1 and L1, and apply the Feldman-Cousins procedure to it \cite{Feldman:1997qc}, effectively increasing the $CR$, thus reducing the distance reach. 

In each configuration, we compute the mean and standard deviation of the distance reaches {over different detector times and frequencies, which accounts for the varying \psd}. Among all configurations, we take the maximum distance reach at each chirp mass. We also require that the time-frequency track of each candidate within $\Tpm$ differs by no more than one frequency bin from the time-frequency track that the signal would follow at 3.5\pn: 

\begin{equation}
    |\fgw(t)-f_{\rm3.5PN}(t)| \leq \frac{1}{\TFFT},
    \label{eqn:fgw0PN3.5PN}
\end{equation}
where $f_{\rm3.5PN}(t)$ is given by Eq. 5.258 in \cite{maggiore2008gravitational}. This criterion has a bigger impact on the allowed candidates to contribute to the upper limits for asymmetric mass-ratio systems, since the mass ratio enters at 1\pn. 

Our search is sensitive to ultra-compact objects inspiraling for which $\Mc\in [10^{-5},10^{-2}]\,\msun$,  and whose time-frequency evolution follow \cref{eqn:powlaws}. Thus, we provide in \cref{fig:equal-mass-dist-const} constraints on the distance reach $\dnf$ for equal-mass systems that are independent of {the nature of the compact objects and of} the binary formation model. From \cref{eqn:dmax}, a power-law behavior of $\dnf\propto \mathcal{M}^{5/4}$ is expected. Fitting the curve, we find a power-law index of 1.24, within 1\% from the theoretical value.

The procedure used to calculate $\dnf$ has been validated through injection studies in both O3 \cite{Miller:2024jpo} and \OFourA (see \cref{app:ul}), in which we have tested performance across different times and frequency bands at fixed chirp mass. Its robustness to variations in curvature within the time–frequency plane is further supported by previous injection campaigns targeting rapidly spinning down isolated \nss in O2 \cite{Miller:2018rbg,LIGOScientific:2018urg}. In those studies, the algorithm tracked signals following distinct power-law time–frequency evolutions with sensitivities consistent with \cref{eqn:dmax}. Together, these results demonstrate that the \gfh performs reliably across a broad range of chirping signals.

\begin{figure}[ht!]
     \begin{center}
\includegraphics[width=0.48\textwidth]{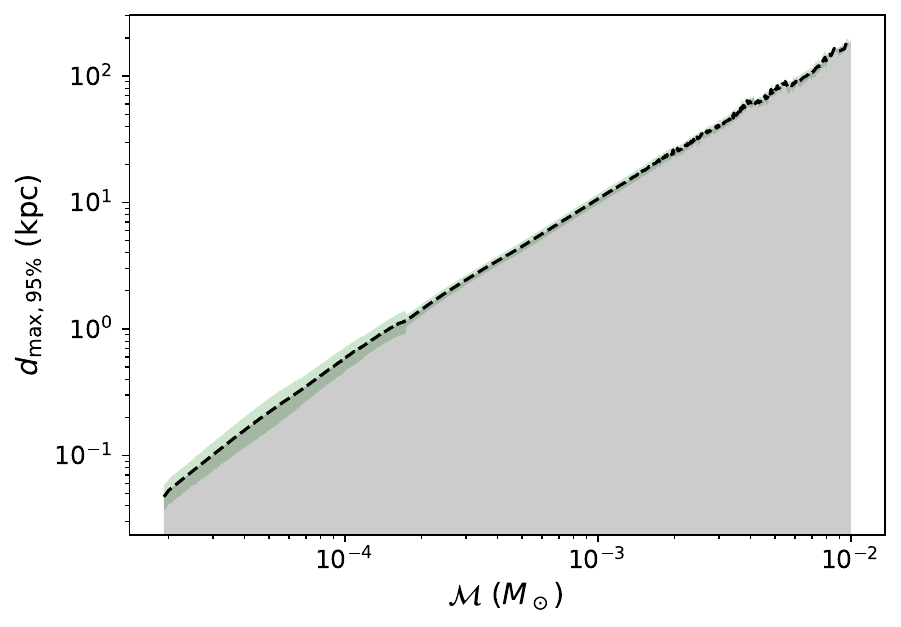}
    \end{center}
    \caption[]{\textbf{Model-independent distance reach constraints for equal-mass inspiraling compact objects}. We show the distance reach at which 95\% of signals would be recovered for equal-mass systems. The green shaded region denotes one standard deviation uncertainty on the distance reach, while the gray-shaded region indicates excluded distances. These constraints follow the expected power-law of $\dnf\propto \Mc^{5/4}$.
    }%
 \label{fig:equal-mass-dist-const}
\end{figure}

\subsection{Computing upper limits on merger rate density}\label{subsec:comp-lim-rate}

{We set upper limits on the merger rate density in a different way than what is done to obtain $\dnf$. This is because $\dnf$ is calculated based on the returned coincident candidates in the search. We set upper limits on rate density not via $\dnf$ to account for the differences between our analysis and those in matched-filter sub-solar mass searches \cite{LIGOScientific:2019kan,LIGOScientific:2021job,Phukon:2021cus,Nitz:2021vqh,Nitz:2021mzz,LIGOScientific:2022hai,Nitz:2022ltl}.

We note that, in standard \ssm searches \cite{LIGOScientific:2019kan,LIGOScientific:2021job,Phukon:2021cus,Nitz:2021vqh,Nitz:2021mzz,LIGOScientific:2022hai,Nitz:2022ltl}, the loudest event across the whole parameter space is used to set upper limits on rate density, which follows the formalism in \cite{Brady:2004gt,Fairhurst:2007qj,Biswas:2007ni}. These are \emph{globally} conservative upper limits. }

{However, our search differs because the detection statistic (the
$CR$) is updated throughout the follow-up procedure. Additionally, the coincident candidates have all been shown to be due to noise disturbances in the follow-up (see \cref{app:final-search} and the figures therein for details on these $CR$s.) In essence, the threshold that we set on our statistic, $\crthr=7$, sets the sensitivity floor of the search: signals with $CR>\crthr$ enter the follow-up stage and can be detected; signals with $CR<\crthr$ are lost. Thus, we can compute the upper limits on rate density by noting that the probability density function of a certain value of the $CR$, given that it originates from a signal at a distance $r$ away, is}

\begin{equation}
     p(CR\mid r) = \frac{1}{\sqrt{2\pi}} e^{-(CR - (D/r)^2)^2/2},
     \label{eqn:pdfCR}
\end{equation}
where $D$ is the collection of prefactors that denote the distance away we could detect a signal as a function of the chirp mass, the frequencies covered by the signal and our analysis parameters $\TFFT,\Tobs$:

\begin{align}
    D&=1.41\left(\frac{G \mathcal{M}}{c^2}\right)^{5/3}\left(\frac{\pi}{c}\right)^{2/3} \frac{T_{\rm FFT}}{{T_{\rm PM}}^{1/2}} \nn\\ &\times\left(\sum_x^N \frac{f^{4/3}_{\text{GW},x}}{S_n(f_{\text{GW},x})}\right)^{1/2}  \left(\frac{p_0(1-p_0)}{Np^2_1}\right)^{-1/4}\label{eqn:D}
\end{align}
See \cref{app:ul} for the full equation and its validation against injections.
$S_n$ is the noise \psd of the H1 or L1, $N=\Tobs/\TFFT$, and $p_0$ and $p_1$ are given in \cref{eqn:p0,eqn:p1}. 
The integral of \cref{eqn:pdfCR} gives the efficiency $\epsilon(r)$:

\begin{equation}
    \epsilon(r) = P(CR > \crthr \mid r) = \frac{1}{2} \, \text{erfc} \left( \frac{\crthr - \left(\frac{D}{r}\right)^2}{\sqrt{2}} \right)
    \label{eqn:eff_D}
\end{equation}
Then, the co-moving spacetime volume $\avgVT$ can be approximated using Laplace's method (see \cref{app:ratecalc} for more details) :

\begin{align}
     \avgVT &= \Tobs \int_{0}^{\infty} 4 \pi r^{2} \, \epsilon(r) \, dr  \nn \\
     &\simeq\Tobs\frac{4}{3}\pi \left(\frac{D}{\sqrt{\crthr}}\right)^3
     \label{eqn:intvt}
\end{align}

Assuming that the event rate for inspiraling ultra-compact objects is Poissonian, consistent with other searches \cite{Phukon:2021cus,LIGOScientific:2021job,LIGOScientific:2022hai}, we can then calculate the upper limits on the rate density at a chosen confidence level $\alpha=0.9$:
\begin{equation}
    \mathcal{R}_{\rm 90\%} = \frac{2.303}{\avgVT}.\label{eqn:R90}
\end{equation}
We show our upper limits on $ \mathcal{R}_{\rm 90\%}$ in \cref{fig:rate-dens-constraints} for equal-mass and asymmetric mass-ratio systems. The rate density follows a power law of approximately $\mathcal{R} \propto D^{-3} \propto {\cal M}^{-15/4}$. From fitting the curve, we find the power-law index to be --3.66, a 2\% difference.

\begin{figure*}[ht!]
     \begin{center}
        \subfigure[ ]{%
            \label{fig:equal-rate}
            \includegraphics[width=0.49\textwidth]{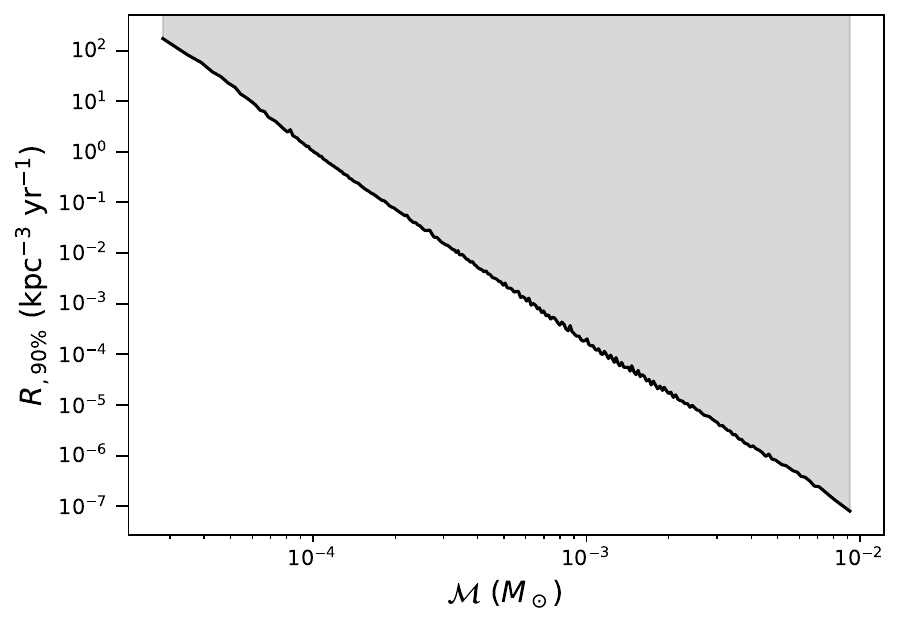}
        }%
        \subfigure[]{%
           \label{fig:asymm-rate}
           \includegraphics[width=0.5\textwidth]{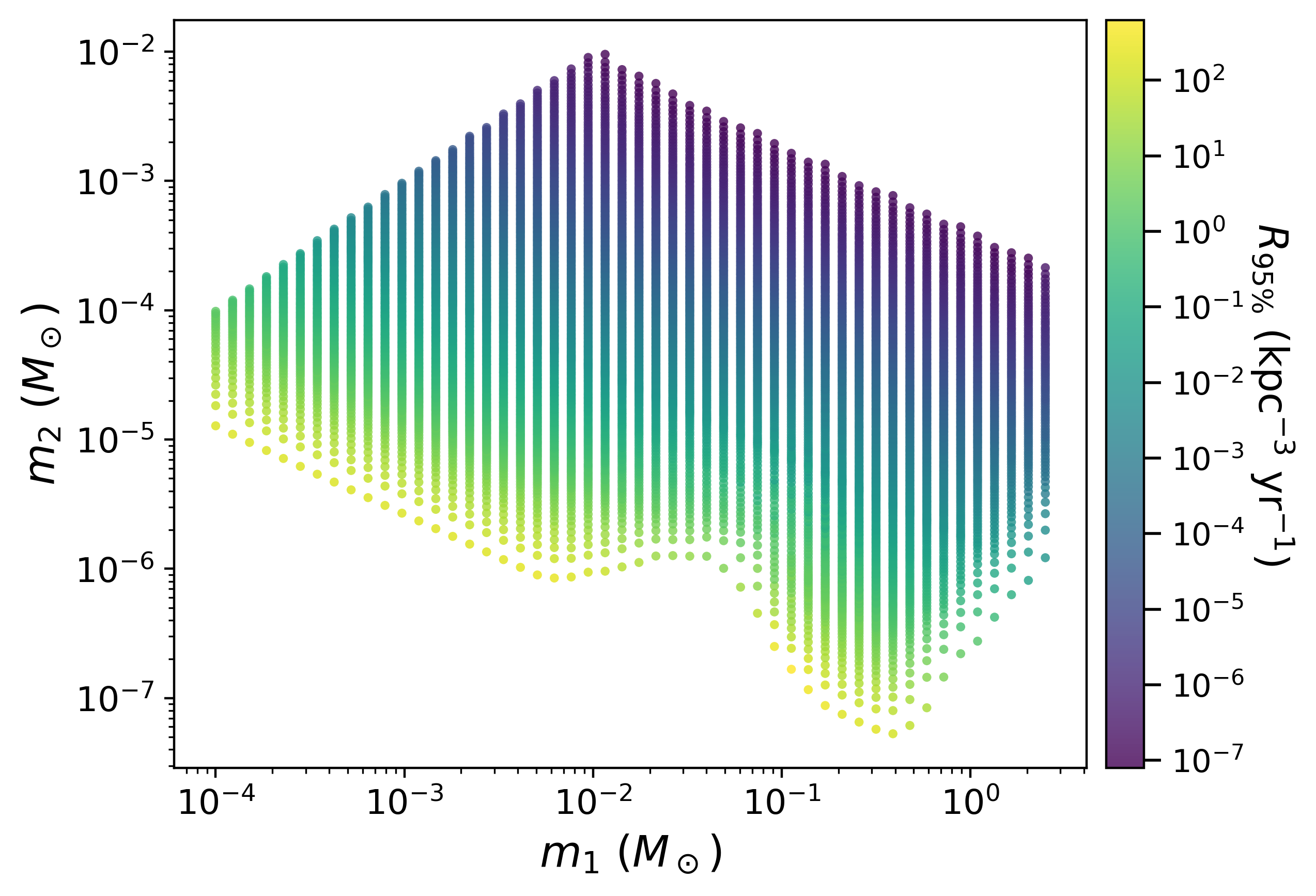}
        }\\ 
    \end{center}
    \caption[]{\textbf{Model-independent constraints on rate density at the 90\% confidence-level for equal-mass (left) and asymmetric mass-ratio inspiraling compact objects (right)}. These constraints are derived following the procedure outlined in \cref{subsec:comp-lim-rate}, and obey the expected power law of $\mathcal{R}\propto\Mc^{-15/4}$. 
    }%
   \label{fig:rate-dens-constraints}
\end{figure*}

\subsection{{Model-dependent constraints on \pbhs}}\label{subsec:dep}

{Translating these constraints into limits on the \pbh abundance is subtle and highly model-dependent.  Indeed, several binary formation channels have been proposed and each of them depend on the \pbh mass distribution and subject to multiple astrophysical uncertainties.  Following state-of-the art rate prescriptions~\cite{Raidal:2024bmm}, early-universe two-body binaries are typically the dominant binary formation channel for the masses and \dm fraction relevant for this work.  Their merger rate densities $\mathcal{R}^{\rm cos}_{\rm prim}$ are given by }  
\begin{align}
        \mathcal{R}^{\rm cos}_{\rm prim}  &\approx   {1.6 \times 10^{-12}}{\text{ \invkpccubedyr}} \ftilde^{53/37} \nonumber \\  &\times \left(\frac{m_1 + m_2}{\,\msun}\right)^{-32/37}\left[\frac{m_1 m_2}{(m_1+m_2)^2}\right]^{-34/37},  \label{eq:cosmomerg}
\end{align}
where we define an effective parameter $\ftilde$ as:

\begin{equation}
\ftilde \equiv \fpbh \left[\fsup\fmone\Delta \ln m_1\fmtwo\Delta \ln m_2\right]^{37/53},
\label{eqn:ftilde}
\end{equation}
where {$f(\ln m)$ is the \pbh mass probability density function, normalized such that $\int f(\ln m)d\ln m =1$. $\fpbh$ is the total fraction of \dm made of \pbhs and $\fsup$ is}
{a merger rate suppression} factor ($=1$ if no suppression) that accounts for {the various mechanisms affecting the binary orbital properties throughout the history of the Universe~\cite{Raidal:2018bbj}, changing their merger time or destroying them\footnote{The fraction of \pbhs that originally form in binaries versus as isolated systems is accounted for in the derivation of \cref{eq:cosmomerg}.}.}
{Constraining 
$\ftilde$ eliminates the dominant sources of uncertainty, namely those arising from the suppression factor and the mass distribution.}  {In the following, we will identify conditions on the mass distributions for which the suppression is minimal, which is required to lead to relevant constraints.}

{Compared to standard \cbc searches, the distance is limited to our galactic environment. One therefore has to translate cosmological merger rates into galactic rates, taking into account the galactic \dmh density profile. }
{If the galactic \dmh density was the one at the Sun's location, $\rho_{\rm DM}\simeq10^{16} \,\msun {\rm Mpc}^{-3}$ \cite{Weber:2009pt}, the merger rates would be enhanced to $\mathcal{R} = 3.3 \times 10^5 \mathcal{R}_{\rm prim}^{\rm cos}$~\cite{Miller:2020kmv}.  However, with \OFourA data, we can probe source distances comparable to the distance to the Galactic Center from Earth. We therefore have modified the rates by a factor $F(d)$ that accounts for the integrated \dmh density profile centered on the sun location at $8.2$ kpc from the galactic center, as described in \cref{app:cosmo-rates}.}
Thus, we rewrite \cref{eq:cosmomerg} in both the equal-mass case
\begin{align}
\mathcal{R} =& 1.04 \times 10^{-6}\, \mathrm{kpc}^{-3} \mathrm{yr}^{-1} F(d) \left(\frac{\mpbh}{\,\msun}\right)^{-32/37}\ftilde^{53/37}~,
\label{eqn:rate-equal}
\end{align}
where $\mpbh=2^{1/5}\Mc$, and the asymmetric mass-ratio case
 \begin{align}
\mathcal{R} &= 5.28 \times 10^{-7}\, \mathrm{kpc}^{-3} \mathrm{yr}^{-1} \nn \\ &\times F(d) \left(\frac{m_1}{\,\msun}\right)^{-32/37} \left(\frac{m_2}{m_1}\right)^{-34/37} \ftilde^{53/37}~,
\label{eqn:rate_asymm}
\end{align}
defining $m_2<m_1$.

\begin{figure}[ht!]
     \begin{center}
\includegraphics[width=0.49\textwidth]{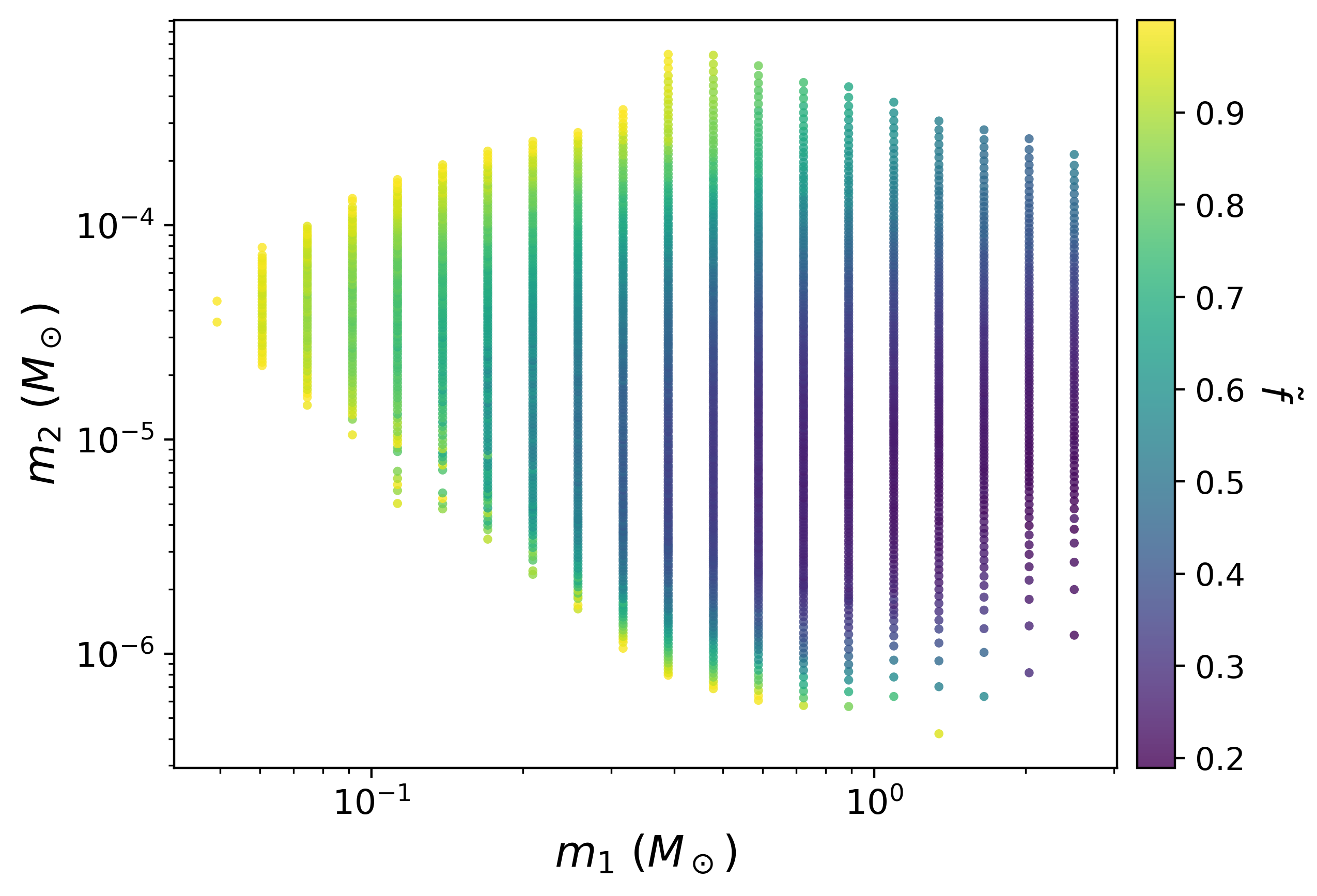}
%
    \end{center}
    \caption{\textbf{Model-dependent constraints on asymmetric mass-ratio inspiraling \pbhs}. We interpret the rate densities in \cref{fig:asymm-rate} as arising from \pbhs, and constrain the effective parameter $\ftilde$ as a function of asymmetric mass-ratio \pbhs using \cref{eqn:rate_asymm}. We restrict the plot only to masses for which $\ftilde<1$, and note that we were unable to constrain $\ftilde<1$ for equal-mass systems.
    }%
      \label{fig:model-dep-constraints}
\end{figure}

In \cref{fig:model-dep-constraints}, we show the constraints on $\ftilde$ in the asymmetric mass-ratio case inferred from \cref{fig:asymm-rate} (we do not find relevant constraints in the equal-mass case).

\subsection{Constraining $\fpbh$ from $\ftilde$}\label{subsec:robust}

{Though we constrain the effective parameter $\tilde f$ down to about 0.2 in a portion of the parameter space for asymmetric mass-ratio binaries, in order to derive a meaningful limit on $f_{\rm PBH}$, given \cref{eq:cosmomerg}, we need to find regimes in which $\fsup \gtrsim 0.04$ or even close to unity.  We identify below some general conditions on the mass distribution for which this is satisfied for asymmetric mass-ratio binaries, and provide further details in \cref{app:cosmo-rates} for equal-mass systems, which remain unconstrained in this analysis.} \tcr{We choose our mass distributions to avoid cases of binary formation suppression ($\fsup\sim \mathcal{O}(1))$; however, to constrain alternative mass functions, the methodology outlined in this subsection can be followed.}
{For asymmetric mass-ratio binaries, we can constrain} 
$f_{\rm PBH}^{53/37} \fmtwo$, under some sufficient conditions:  (i) if there is a peak in the distribution at the mass $m_1$, such that $\fmone\Delta \ln m_1 \approx 1$ and the mean \pbh mass is $\langle m \rangle \approx m_1 $, and (ii) if there is at least a small \dmh fraction in heavy black holes, enough to seed \pbh clusters, such that the binary is unlikely to be perturbed by other \pbhs. 
{In this case, we obtain $\fsup \approx 0.5$ and can constrain $\fmtwo$ in the range $10^{-6} < m_2/\msun < 10^{-4}$, 
if $m_1$ is at the solar mass scale, as expected in the motivated class of mass functions imprinted by the {QCD epoch}~\cite{Byrnes:2018clq}. The most stringent limit on $\fmtwo$ is when one assumes $\fpbh = 1$, but one should note that this would be in conflict with existing limits. We show these limits on $\fmtwo$ assuming $\fpbh=1$ in \cref{fig:money}, which can be compared to existing experiments.}



\section{Conclusions}\label{sec:concl}

\begin{figure*}[ht!]
    \centering
    \includegraphics[width =\linewidth]{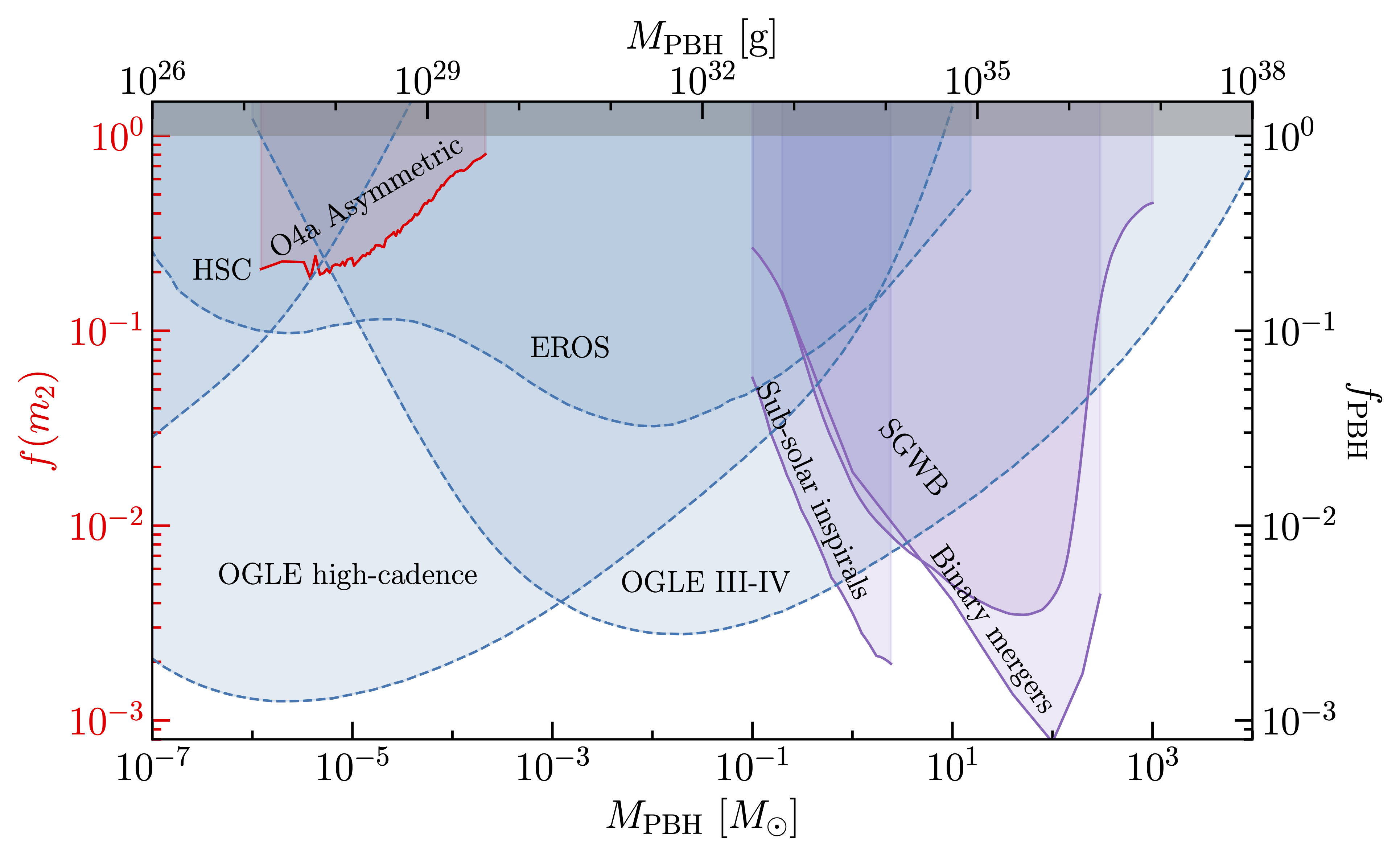}
    \caption{{\textbf{Constraints on \pbhs from this search, in red, and other \gwh and electromagnetic analyses.} {Our limits on the mass function $f(m_2)$ (shorthand for $\fmtwo$) are valid for mass distributions respecting the conditions mentioned in the text for asymmetric mass ratio binaries, assuming $\fpbh=1$. No curve is shown for equal-mass systems, as $\tilde{f} < 1$ is not constrained. We assume $\fmone\Delta\ln m_1\sim 1$ and use the corresponding values of $\ftilde$ in \cref{fig:model-dep-constraints} at $m_1=2.5\,\msun$ to obtain the red line on this plot. We emphasize that these constraints are valid only for the classes of mass functions discussed in \cref{subsec:robust}. If one wishes to produce our constraints for different choices of $\fpbh$, the red curve will be scaled upwards by $\fpbh^{-53/37}$. } Constraints from other probes (microlensing and \gws) are also presented for comparison on $\fpbh$, and are implicitly valid for monochromatic mass functions and subject to astrophysical uncertainties.}
    Purple curves correspond to constraints from previous \gwh searches \cite{Kavanagh:2018ggo,LIGOScientific:2019kan,LIGOScientific:2021job,LVK:2022ydq,Nitz:2022ltl,Wang:2023qgw,Boybeyi:2024mhp,LIGOScientific:2025kry}, while dashed blue curves indicate microlensing constraints \cite{EROS-2:2006ryy,Croon:2020ouk,Mroz:2024wia,Mroz:2024mse} that could weaken significantly due to \pbh clustering \tcr{depending on the assumed \pbh formation model, mass of the cluster and the distributions of \dmh density and velocity} \cite{Gorton:2022fyb,Green:2025dut}. Our constraints on the mass function in red should be referenced to the red left $y-$axis, while all other constraints are directly on $\fpbh$ and should be referenced to the right $y-$axis.}
    \label{fig:money}
\end{figure*}

We have performed a search for planetary-mass, ultra-compact objects using data from the first part of the \lvk fourth observing run. Though we did not find any significant events, we placed upper limits on both the distance reach and merger rate density of equal-mass and asymmetric mass-ratio ultra-compact binaries. Moreover, for certain classes of formation models, we constrain the fraction of \dm that \pbhs could compose for asymmetric mass-ratio systems.
Our work complements \mf searches for \ssm \pbh binaries by considering lower-mass regimes that would require too much computational power to analyze with \mf. 
Our results provide a complementary way to probe planetary-mass \pbh binaries that could form in clusters. In \cref{fig:money}, {we show how our bounds on $\fmtwo$} compare to other observations, assuming $\fpbh=1$. We also find that our constraints are valid for systems with eccentricities as high as 0.84 (see \cref{app:robust} for more details).
Note that each method has its own systematics and assumptions, and it may happen that microlensing constraints weaken due to a better understanding of the galactic rotation curves~\cite{Garcia-Bellido:2024yaz}, which strengthens the argument that we need multiple probes of \pbhs in this mass regime. Moreover, we note that \pbh physics is an evolving field: while we consider particular mass functions and conditions to arrive at \cref{fig:money}, we emphasize that these limits could change with time, and that different assumptions could lead to different constraints. Hence, we release the rate density constraints directly to allow readers to choose their own ways of interpreting our results, along with codes to generate these plots \cite{bradley_j_kavanagh_2019_3538999,miller_2025_15559328,miller_2025_15512455,LIGO_P2500686}.

As we look forward to future ground- and space-based \gwh \ifos that will probe even lower frequencies than accessible now, the impact of eccentricity 
on the inspiral signal 
will become even more significant. Waveforms to handle eccentricity \cite{Ramos-Buades:2021adz,Liu:2023ldr,Gamba:2024cvy,Nagar:2024dzj,Gamboa:2024hli,Bhaumik:2024cec,Morras:2025xfu,Morras:2025nlp,Planas:2025feq,Huez:2025npe} and \dmh effects \cite{Bertone:2019irm,Kavanagh:2020cfn,Cole:2022yzw,Kavanagh:2024lgq,Karydas:2024fcn,Dyson:2025dlj,Mitra:2025tag} in \mf analyses are currently under development, but these effects will not impact semi-coherent analyses, such as the \gfh, as much. Our analysis thus allows us to be sensitive to a range of rich physics while maintaining computational efficacy.

\appendix

\section{Details on final steps of the search}\label{app:final-search}

We select the top 1\% of candidates in every Hough map that is created. Each candidate is defined to have the following parameters (1) the signal frequency at a reference time, (2) the chirp mass, and (3) the detection statistic. The analysis of each detector's data is performed separately, meaning that we can look for \emph{coincident candidates}, i.e. candidates with similar enough parameters present at the same times in both detectors. We define ``close enough'' to be three bins away:

\begin{equation}
    \text{dist} = \sqrt{\left(\frac{k_{\rm LHO} - k_{\rm LLO}}{\delta k}\right)^2 + \left(\frac{z_{0,\rm LHO} - z_{0,\rm LLO} }{\delta z_0}\right)^2}
    \label{eqn:dbin}
\end{equation}
where $\delta k$ and $\delta z_0$ are the bin sizes in each of the coordinates in the Hough map. Note that we create the Hough maps in the transformed coordinates $z=1/f^{8/3}$ and $k$, not $f_0$ and $\Mc$ -- see \cite{Miller:2018rbg} for details on this transformation. $\delta k$ varies as a function of $\TFFT$, $k$ ($\Mc$) and $f_0$, while $\delta z_0$ depends solely on $f_0$ and $\TFFT$ \cite{Miller:2018rbg}. 

At this stage, we apply a threshold on the critical ratio, $\crthr\gtrsim 7$, that ensures that we only consider candidates with a false alarm probability of 1\% (accounting for the trials factor) in Gaussian noise. The data are not Gaussian, which means that we tend to keep large noise disturbances as well as potential signals. If we were to have estimated a background for the $CR$ using time-slides or another method traditionally used in compact binary searches, we would have obtained a higher threshold than $\crthr\sim 7$ due to the presence of non-Gaussian noise disturbances. Thus, our threshold is conservative. We show the distribution of candidate above this threshold in \cref{fig:loudest-event}.

The number of coincident candidates that surpass these tests is $\sim 5\times 10^5$. Each of these candidates is subject to a follow-up procedure: the data are demodulated based on the candidate parameters, which would lead to a perfectly monochromatic signal if the demodulation was done correctly. The search is then re-run, which a new $\TFFT$ equal to twice the original one.  All candidates are vetoed in this procedure, since they do not produce $CR$s that exceed those from the first stage of the search, as would be expected for a monochromatic signal with increased $\TFFT$ length.

\section{Obtaining distance reach upper limits}\label{app:ul}

We describe how we set upper limits on the model-independent distance reached as a function of chirp mass.
To do so, we employ a semi-analytic/ data-driven procedure outlined in \cite{Miller:2024jpo}, which uses as input results from our search (detection statistics and detector \psd), and an equation for theoretical sensitivity estimate of our method, as a way to obtain upper limits on the maximum distance away that we could have seen a source as a function of chirp mass, i.e. \cref{fig:equal-mass-dist-const}. 
Such a procedure was designed to avoid extensive injection campaigns to set upper limits and instead use that computing power to follow-up interesting candidates.

Even though each configuration is labeled by a particular chirp mass and starting frequency ($\mconfigi$,$\fconfigsi$), the \pmaps and Hough maps generated can actually probe many different chirp masses and starting frequencies. 
Let us label the space of chirp masses as $\mathcal{M}_k$. 
Within a given Hough map (say $j$) for a particular configuration (say $i$), all instances of $\mathcal{M}_k$ (with any starting frequency $f_l$) are assigned a critical ratio ($CR^{IFO}_{i,j,k,l}$) per interferometer $\rm IFO$. 
We then compute the distance reach using \cref{eqn:dmax} for each detector separately with the following:

\begin{align}
d^\Gamma_{\rm max}&=1.41\left(\frac{G \Mc}{c^2}\right)^{5/3}\left(\frac{\pi}{c}\right)^{2/3} \frac{\TFFT}{{\Tpm}^{1/2}}\left(\frac{p_0(1-p_0)}{Np^2_1}\right)^{-1/4} \nonumber \\ &\times \left(\sum_x^N \frac{f^{4/3}_{\text{GW},x}}{S_n(f_{\text{GW},x})}\right)^{1/2}{\left(CR-\sqrt{2}\erfc^{-1}(2\Gamma)\right)^{-1/2}}.
\label{eqn:dmax}
\end{align} 
where $\Gamma=0.95$ is the fraction of detectable signals in repeated experiments, $N=\Tobs/\TFFT$ is the number of FFTs used to make the \pmap, and $p_0$ is the probability of selecting a peak above  $\thetathr=2.5$, the threshold on equalized power in the \pmap, and:

\begin{align}
    p_0=e^{-\thetathr}-e^{-2\thetathr}+\frac{1}{3}e^{-3\thetathr}
    \label{eqn:p0}
\end{align}
and

\begin{align}
    p_1 &= \thetathr \left( \frac{1}{2} e^{-\thetathr} - \frac{1}{2} e^{-2\thetathr} + \frac{1}{6} e^{-3\thetathr} \right) \nn \\ &+ \frac{1}{4} e^{-2\thetathr} - \frac{1}{9} e^{-3\thetathr}.
    \label{eqn:p1}
\end{align}
both relate to the probability of selecting a peak above the threshold $\thetathr$ in the presence of a weak monochromatic signal \cite{Palomba2025_peakmap}. We note that this equation is strictly valid for signals lasting at least one sidereal day. However, since it is evaluated millions of times for coincident candidates occurring at random times in the run, the ensemble of evaluations effectively samples all sidereal phases.

{In \cref{eqn:dmax}, we must input a value for the $CR$ that accounts for the fact that this equation was derived in the case of Gaussian noise, but our analysis was done in real noise. We first take the maximum of the two $CRs$ returned from each \ifo: $CR_{i,j,k,l}$=max$_{\rm IFO}(CR^{IFO}_{i,j,k,l})$. This is a conservative choice, since it reduces $d^\Gamma_{\rm max}$ with respect to using the average or minimum $CR$.

Then, we employ the Feldman-Cousins (FC) \cite{Feldman:1997qc} approach, in which we assume that the $CR$ follows a Gaussian distribution, 
and map the measured $CR_{i,j,k,l}$ to a positive-definite value using the upper value of Tab. 10 in \cite{Feldman:1997qc} at 95\% confidence. Here, we input $CR_{i,j,k,l}$ into the Feldman-Cousins procedure. In other words, $CR_{\mathrm{FC},i,j,k,l}=FC(CR_{i,j,k,l})$ in \cref{eqn:dmax}, which serves to reduce the distance reach with respect to if we had used just the maximum $CR$ of the coincident candidates. Thus, the meaning of the upper limit is as follows: a real signal would need to have a critical ratio larger than that used in \cref{eqn:dmax} in order to be detectable in 95\% of repeatable experiments, at a particular reference time, chirp mass and frequency.

{Let us call the result of inputting $CR_{\mathrm{FC},i,j,k,l}$ into \cref{eqn:dmax} $d^\Gamma_{i,j,k,l}$. For each combination of $i,j,k,l$ -- which corresponds to each coincident candidate we have in the search, before applying any threshold or follow-ups --, we compute a distance reach at which 95\% of signals in a repeated number of experiments would be recovered, $d^{95\%}_{i,j,k,l}$, which represents, in repeated experiments, the distance at which 95\% of signals with a given chirp mass at a given time during the run would have been detected.

We then take the median of all the distances over the duration of the observing run, i.e. $d^{95\%}_{i,k}$ = median($d^{95\%}_{i,j,k}$) over each set of $\Tobs/\Tpm$ values, leading to one distance reach per configuration per probed chirp mass. These denote the different sensitivities of each configuration to that particular chirp mass, some of which will be more sensitive than others. Therefore, we take the maximum of these distances for each chirp mass as the distance upper limit, i.e. $d_\mathrm{max,95\%,k} = \mathrm{max}(d^{95\%}_{i,k})$, which are shown in \cref{fig:equal-mass-dist-const}.}

The upper limits in \cref{fig:equal-mass-dist-const} are independent of any population or formation model for planetary-mass ultra-compact objects. They indicate that a system with a given chirp mass could have been detected in 95\% of repeatable experiments at a particular distance away from us. 

As shown in \cite{Miller:2018rbg,Miller:2020kmv,KAGRA:2022dwb,Miller:2024jpo}, the use of \cref{eqn:dmax} produces consistent upper limits compared to those that would be obtained by injecting simulated signals. We also provide an injection study done in this dataset for a range of chirp masses in \cref{fig:o4a-new-inj-study}. For each chirp mass, we perform fifty injections over a range of distances away from us, and determine the distance reach at which we recovered at least 95\% of injections above $\crthr=7$ in both detectors. We compare our results to the theoretical distance reach at 95\% efficiency given by \cref{eqn:dmax}. Note that \cref{eqn:dmax} represents population-averaged upper limits; therefore, we must specialize this equation to the specific sources we simulate in order to compare them to the sensitivity obtained through injections. In particular, we account for the fact that we have simulated fifty sources with (1) specific choices of inclination and polarization angles, and sky position, and (2) durations that are much less than a sidereal day, both of which affect \cref{eqn:dmax}. This per-source factor $\mathcal{C}_s$, where $s$ denotes the source number, must be divided out of \cref{eqn:dmax} to specialize it to a particular source \cite{Palomba2025_peakmap}, as discussed previously in the appendices of \cite{KAGRA:2022dwb,Miller:2023qph}:

\begin{equation}
\mathcal{C}_s = \sqrt{\frac{\mathcal{S}^2_{\alpha, \delta, \psi, \cos\iota}}{\mathcal{S}^2_t}},
\end{equation}
where
\begin{equation}
\mathcal{S}^2_{\alpha, \delta, \psi, \cos\iota} = \langle (F_+A_+ + F_\times A_\times)^2 \rangle_{\alpha, \delta, \psi, \cos\iota} \approx \frac{4}{25}
\end{equation}
is the factor obtained normally in the derivation of \cref{eqn:dmax} by averaging the induced strain over one sidereal day, sky position $\alpha,\delta$, cosine of the inclination angle $\cos\iota$, and the polarization angle $\psi$. Additionally, $A_+$ and $A_\times$ are the plus and cross polarizations of the \gwh, given by:

\begin{align}
A_+ &=\frac{1 + \cos^2 \iota}{2}, \\
A_\times &= \cos\iota.
\end{align}
The factor $\mathcal{S}^2_t$ is obtained at the specific parameters of the source:
\begin{equation}
\mathcal{S}^2_t = \langle F^2_{+}\rangle_t A_+^2  + \left\langle F^2_{\times}\right\rangle_t A_\times^2,
\end{equation}
where $\psi$ has been averaged out, and the average over time $t$ is taken only over the duration of the source, and
\begin{align}
F_{+}(t) &= a(t) \cos 2\psi + b(t) \sin 2\psi,  \\
F_{\times}(t) &= b(t) \cos 2\psi - a(t) \sin 2\psi,
\end{align}
are the time-varying beam pattern functions. The functions $a(t)$ and $b(t)$ are given in \cite{jaranowski1998data}.

We calculate $\mathcal{C}=\rm mean(\mathcal{C}_s)$ as an average over the 50 source-specific factors $\mathcal{C}_s$, and then divide \cref{eqn:dmax} by $\mathcal{C}$ to plot the dark blue curve in \cref{fig:o4a-new-inj-study}. Furthermore, we shade the $\pm 1\sigma$ around the dark blue curve, which encapsulates the range of possible $\mathcal{C}$, i.e. $\sigma=\rm std(\mathcal{C}_s)$. Finally, the upper limits that we quote in \cref{fig:equal-mass-dist-const} rely on the application of the Feldman-Cousins procedure, so we reevaluate the theoretical curve at $CR_{\rm FC}=FC(\crthr=7)=8.96$ for comparison. We see that the injections lie comfortably within the shaded area and are consistent with both the source-specific theoretical curve and the Feldman-Cousins curve.


\begin{figure}[ht!]
     \begin{center}
%
           \includegraphics[width=0.49\textwidth]{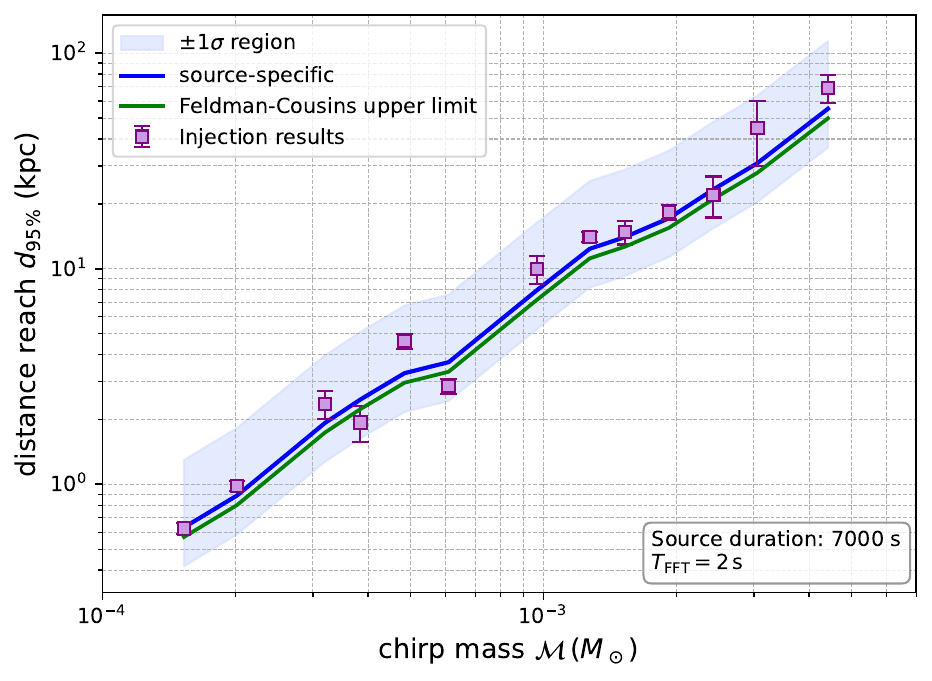}
%
    \end{center}
    \caption{\textbf{Comparison of upper limits on distance reach calculated through injections and with the formula given in \cref{eqn:dmax}}. The blue line indicates the theoretical expectation for the distance reach at 95\% confidence, adjusted by the source-specific factor $\mathcal{C}$ (see text for details). The shaded region denotes the $\pm1\sigma$ uncertainty on the distance reach based on the range of source-specific $\mathcal{C}_s$ values. The magenta points indicate the derived distance reach at which 95\% of injections are recovered in our simulations. The green line indicates the upper limit that we would actually quote based on the Feldman-Cousins procedure. Our results show that the injection studies agree well with the theoretical blue and green curves and fall well within the shaded region. For each set of fifty injections, at each chirp mass, performed at different distance reaches to derive these results, we analyze different frequency bands and different times across the \OFourA dataset. }
      \label{fig:o4a-new-inj-study}
\end{figure}

\begin{figure}[ht!]
    \centering
    \includegraphics[width=\linewidth]{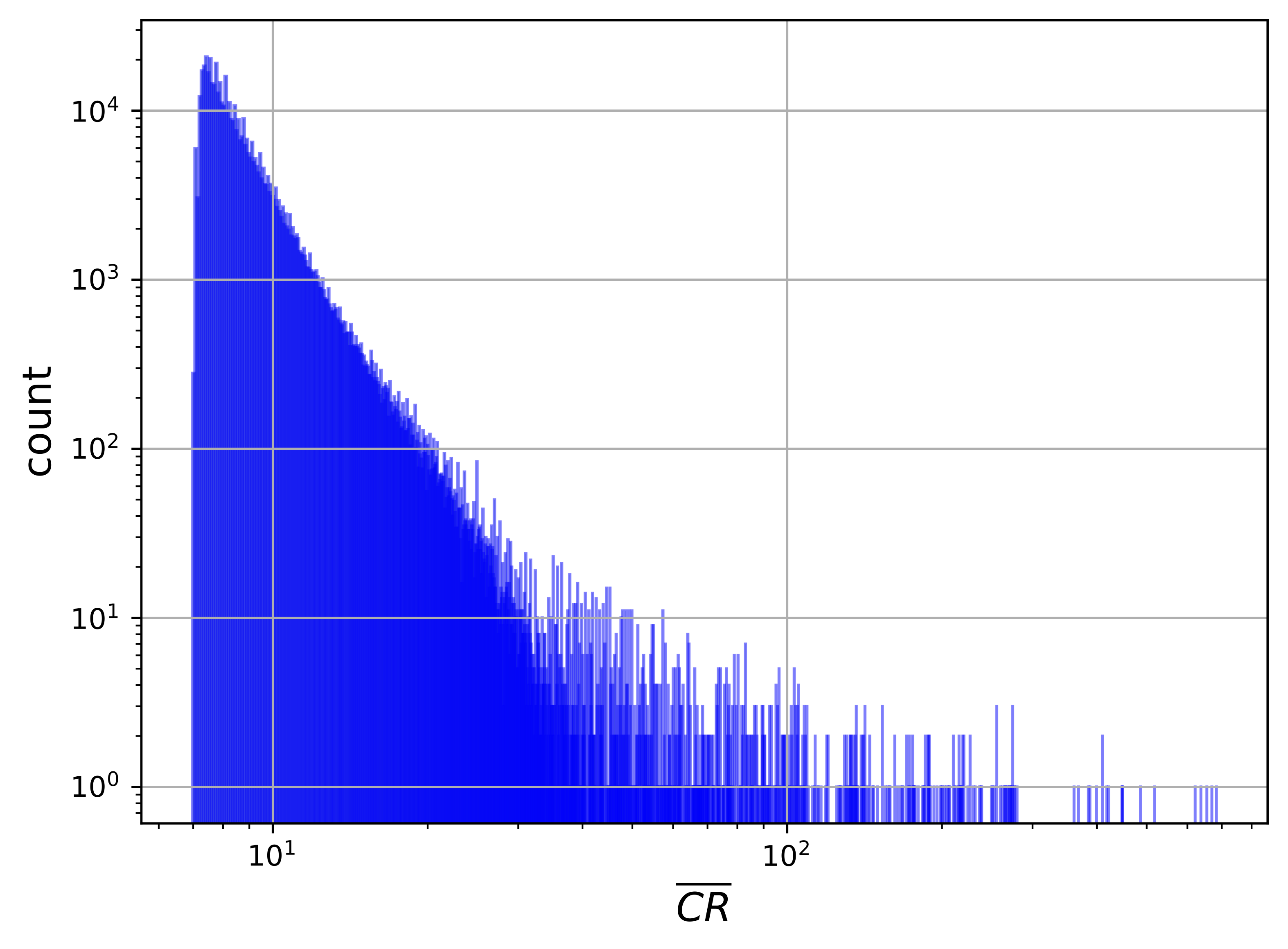}
    \caption{\textbf{The critical ratios of coincident candidates returned in our search whose $\overline{CR}>\crthr=7$}. Candidates with $CR<\crthr$ are not shown. $\overline{CR}$ is averaged from candidates returned in both H1 and L1, and corresponds to the candidates that we followed up, all of which were vetoed. In order to compute upper limits on distance reach, we use these $CR$s in \cref{eqn:dmax} and apply the Feldman-Cousins procedure \cite{Feldman:1997qc} that effectively increases $CR$ to ensure  95\% confidence-level coverage. In other words, real \gwh signals would have had to generate $CR$s above the ones in this figure at a given chirp mass, time and frequency to have been detected by our search 95\% of the time.}
    \label{fig:loudest-event}
\end{figure}

\section{Obtaining rate density upper limits}\label{app:ratecalc}
{
For a fixed chirp mass and frequency range, the only parameter that affects the efficiency of the search is the \gwh amplitude -- or, equivalently, the source distance -- from the \ifos, so the space-time volume $\avgVT$ to which we are sensitive can be written as}

\begin{equation}
    \avgVT = \Tobs \int_{0}^{\infty} 4 \pi r^{2} \, \epsilon(r) \, dr ,
    \label{eqn:intvt-app}
\end{equation}
{$r$ is an arbitrary distance and $\epsilon(r)$ is the efficiency curve, given by \cref{eqn:eff_D}. To compute $\avgVT$, we can use Laplace's method, which allows us to solve this integral analytically by considering its asymptotic behavior. We define

\begin{align}
   x\equiv\frac{\crthr - \left(\frac{D}{r}\right)^2}{\sqrt{2}},
\end{align}
where $D$ is given by \cref{eqn:D} and $\crthr=7$ is the threshold on the critical ratio that we use to determine which candidates to follow up,
and expand linearly around $r_\star= D/\sqrt{\crthr}$, which is the point at which $x=0$:

\begin{align}
    x(r)&\simeq x(r=r_\star) +\left.\frac{dx}{dr}\right|_{r = r_\star}(r-r_{\star}), \nn \\
    &\simeq \frac{\sqrt{2}\crthr^{3/2}}{D}(r-r_\star),
\end{align}
and thus
\begin{equation}
    \epsilon(r)\simeq \frac{1}{2}\erfc\left(\frac{\sqrt{2}\crthr}{D}(r-r_\star)\right).
\end{equation}
To leading order, we can replace $\epsilon(r)$ by a step function at the transition point $r_\star$, where the inverse error function $\erfc$ starts to rapidly decrease for values $r>r_\star$.
\begin{align}
    \avgVT &\simeq \Tobs \int_{0}^{r_{\star}} 4 \pi r^{2} \, dr , \\
    &\simeq \Tobs \frac{4}{3}\pi \left(\frac{D}{\sqrt{\crthr}}\right)^3.
\end{align}
Suppose that we would like to check what the next-to-leading correction would be, to ensure that it is small with respect to what we quote in \cref{eqn:intvt}. To do this, we can change variables such that $u=a(r-r_\star)$, where $a=\frac{\sqrt{2}\crthr^{3/2}}{D}$ and write the integral as

\begin{align}
    \avgVT&= \Tobs4\pi\int_{0}^{\infty} \epsilon(r) \, r^2 \, dr \nn \\ &\approx  \Tobs4\pi\frac{1}{a} \int_{-ar_{\star}}^{\infty} \frac{1}{2} \, \erfc(u) \left( r_{\star} + \frac{u}{a} \right)^2 du.
    \label{eq:integral_approximation}
\end{align}
Splitting $\erfc(u)$ into a step-function piece and a correction piece:

\begin{equation}
\frac{1}{2} \, \text{erfc}(u) = H(-u) + s(u),
\end{equation}
where
\begin{equation}
s(u) \equiv \frac{1}{2} \left( \text{erfc}(u) - 2H(-u) \right) = 
\begin{cases} 
\frac{1}{2} (\text{erfc}(u) - 2), & u < 0, \\
\frac{1}{2} \text{erfc}(u), & u > 0,
\end{cases}
\end{equation}
and noting that $s(u)$ has the following properties:

\begin{equation}
    \int_{-\infty}^{\infty} s(u) \, du = 0, \quad \int_{-\infty}^{\infty} u \, s(u) \, du = \frac{1}{4},
    \label{eq:identities}
\end{equation}
we can calculate the next-to-leading-order piece of $\avgVT$:

\begin{equation}
    \Delta \avgVT = 4\pi \Tobs \frac{1}{a}\int_{-\infty}^{\infty} \frac{1}{2} \, s(u) \left( r_{\star} + \frac{u}{a} \right)^2 du.
\end{equation}
Here, we have extended the lower bound from $-ar_\star$ to $-\infty$ because the integral takes on the largest values around $r=r_\star$
Using \cref{eq:identities} and dropping the $u^2$ term, we arrive at:

\begin{align}
    \Delta \avgVT &= 4\pi \Tobs \frac{r_{\star}}{2a^2},\nn \\
    &= 4\pi \Tobs \frac{D^3}{4\crthr^{7/2}}.
\end{align}
Finally, we can express the leading- and next-to-leading-order contributions as:

\begin{align}
    \avgVT_{\rm tot} &\simeq \avgVT +\Delta \avgVT \nn \\
    &\simeq\Tobs\frac{4}{3}\pi \left(\frac{D}{\sqrt{\crthr}}\right)^3\left[1+\frac{3}{4\crthr^2}+\dots\right]
\end{align}
Given that we use $\crthr=7$ in this paper, the next-to-leading-order term is of $\mathcal{O}(10^{-2})$ and can thus be neglected, allowing us to safely set $\avgVT_{\rm tot}=\avgVT$.

{From $\avgVT$, we can compute the rate density constraint by assuming that the event rate for inspiraling planetary-mass ultra-compact objects follows a Poissonian distribution, which is consistent with our expectations of how \bbh objects inspiral and merge.  The rate density equation can be derived independently of \gwh data analysis: we only need to consider inspirals as a Poisson process with a mean number of expected events of
$\lambda = \mathcal{R}\avgVT$. In a Poisson process, the probability that we observe zero events $N$ is:}

\begin{equation}
    P(N=0|\lambda) = e^{-\lambda}
\end{equation}
{To obtain an upper limit at a given confidence level $\alpha$, we simply need to set:}

\begin{equation}
    P(N=0|\lambda) = 1-\alpha = e^{-\lambda}
\end{equation}
By setting $\alpha=0.9$ and solving for $\mathcal{R}$, we obtain \cref{eqn:R90}. 

\section{From cosmological to galactic rates}\label{app:cosmo-rates}

In this section, we describe the methodology to obtain galactic merger rates as a function of the distance reach $D$ outputted by the search.  The expected rate is obtained from the cosmological merger rate densities, multiplied by the probe volume $\frac{4}{3} \pi D^3$ and by the ratio between the averaged \dmh density in this volume and the cosmological density. We use a standard Navarro-Frenk-White profile,
\begin{equation}
\rho(r) = \frac{\rho_0}{ \frac{r}{ R_{\rm s} }  \left( 1+  \frac{r}{ R_{\rm s} }\right)^2}~,
\end{equation}
with $R_{\rm s} = 21.5$ kpc and $\rho_0 = 8 \times 10^6 \,\msun {\rm kpc}^{-3}$. Using another type of profile, such as a Einasto profile, only changes our results marginally.  The enclosed \dmh mass $M_{\rm DM} (D)$ is obtained by integrating the profile centered on the Sun's location, at $R_\odot = 8.2$ kpc from the galactic center, to get
\begin{align}
& M_{\rm DM} ({D}) = 2 \pi \int_0^\pi \int_0^{D}   r^2 \nonumber \\ & \times \rho\left(\sqrt{r^2 + R_\odot^2 - 2 r R_\odot \cos(\phi)}\right) \sin(\phi) {\rm d} r {\rm d} \phi \,. 
\end{align}
We then compute the factor $F(d)$ as the ratio between the averaged density at distance $d$ and the density at the sun location.  The obtained value of $F(d)$ is shown in \cref{fig:fofd}.  As expected, it is close to one for short distances compared to the galaxy scale.  It shows a slight peak at $d\sim 8$ kpc that corresponds to the distance of the galactic center where \dmh density is strongly enhanced.  But we note that the rate enhancement is small because the search is sensitive to all directions of the sky and not only the galactic center.  At larger distances, $F(d) $ decreases and above 50 kpc, it decreases like $1/d^3$ as expected given that the \dmh  density decays and the total mass probed does not vary much.  We then obtain a constant rate value compatible with the one calculated in~\cite{Miller:2020kmv}, $R^{\rm gal} \approx 2 \times 10^{-8}  \times \mathcal{R}_{\rm cos}^{\rm prim}\, {\rm Gpc}^3$.

\begin{figure}[ht!]
    \centering
    \includegraphics[width =\linewidth]{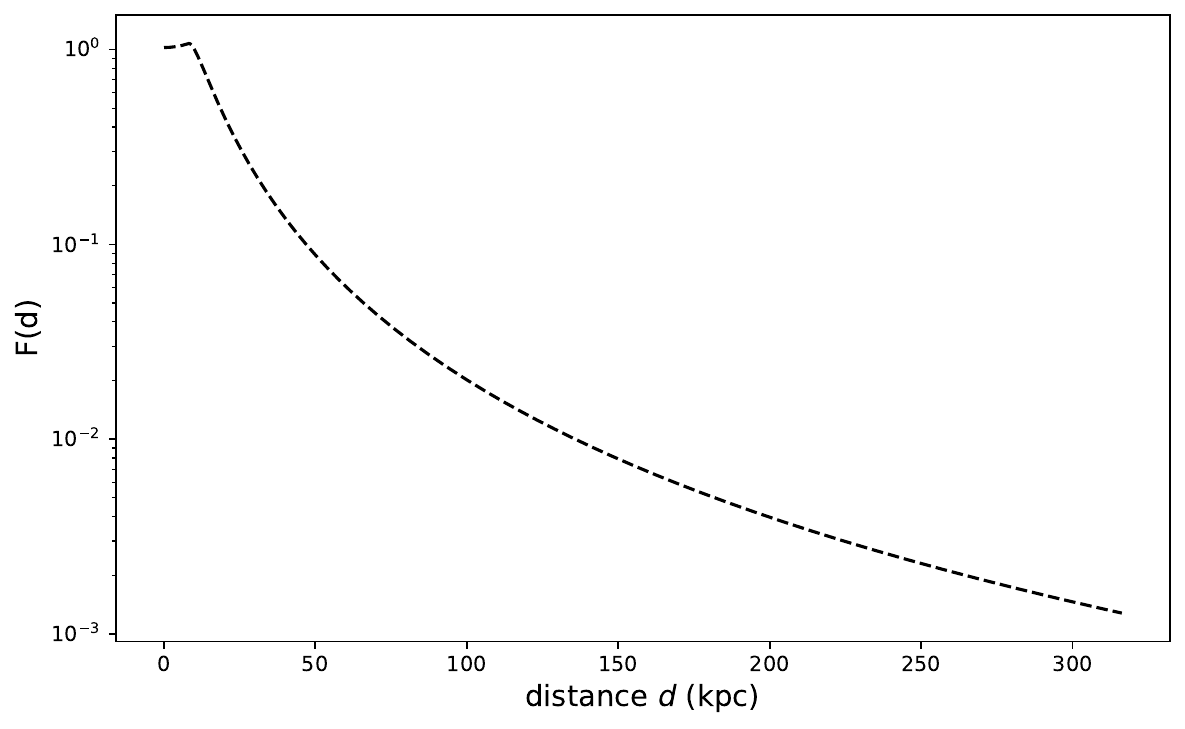}
    \caption{\textbf{How the factor $F(d)$ falls off as as a function of distance from Earth}. $F(d)$ encodes the ratio of the averaged \dmh density at a distance $d$ compared to that at the Sun's position.}
    \label{fig:fofd}
\end{figure}

\section{Evaluating robustness of constraints}\label{app:robust}

\subsection{Eccentricity}

Our constraints have implicitly assumed that \gws from inspiraling compact objects follow the time-frequency evolution $f_{3.5\pn}(t)$. It is, however, worth discussing whether additional physics could affect our ability to set upper limits on \pbhs. We consider to what extent  
binary eccentricity
would alter our conclusions. \tcb{Early-universe \pbhs have merger timescales of the order of the age of the universe; therefore, they are expected to radiate away their eccentricity by the time that they enter the \lvk frequency band \cite{Musco:2012au,Nakamura:1997sm,Ioka:1998nz}. However, \pbh binaries formed by dynamical capture due to \gwh braking in \dmh halos in the late universe would retain some eccentricity, depending on the velocity dispersion and the halo mass, with higher relative velocities (that result from larger halos masses) leading to tighter binaries and larger eccentricities \cite{Cholis:2016kqi,Wang:2021qsu}.
Therefore, the detection of eccentricity could indicate a particular type of binary formation.}

{To demonstrate our sensitivity to eccentric systems, we require that the frequency shift induced by eccentricity 
would be confined to one frequency bin for the total signal duration. In other words, we cannot distinguish between $f_{3.5\pn}(t)$ and the signal time-frequency evolution with eccentricity.
{The key quantity here is the time-frequency track:
eccentricity does not cause frequency fluctuations throughout the signal's time-frequency track by more than one frequency bin.} In other words, if \cite{maggiore2008gravitational,Miller:2024khl}

\begin{equation}
\fdotgw g(\ecc)\TFFT \leq \frac{1}{\TFFT}
\label{eqn:ecc-fdot-limit}
\end{equation}
we would not miss eccentric systems. The function that characterizes the eccentricity $\ecc$ is given by \cite{maggiore2008gravitational}:

\begin{equation}\label{eqn:g_of_ecc}
    g(\ecc) = \left(1-\ecc^2\right)^{-7/2} \left(1 + \frac{73}{24}\ecc^2 + \frac{37}{96}\ecc^4\right).
\end{equation}
Plugging \cref{eqn:fdot_chirp} into \cref{eqn:ecc-fdot-limit} and solving for $g(\ecc)$, we obtain:

\begin{equation}
g(\ecc)\lesssim 8000 \left(\frac{1\,\mathrm{s}}{\TFFT}\right)^2 
\left( \frac{10^{-3} \,\msun}{\mathcal{M}} \right)^{5/3}
\left( \frac{100\,\mathrm{Hz}}{\fgw} \right)^{11/3}.
\label{eqn:max-g-ecc}
\end{equation}
Solving for the eccentricity is analytically unfeasible, so we numerically compute the maximum eccentricity $\ecc_{\rm max}$ to which each configuration is sensitive. Given that each configuration covers a wide range of $\fgw$ and $\Mc$, we calculate the median $\ecc_{\rm max}$ by sampling over the range of $\fgw,\Mc$ each configuration could detect. We plot these median $\ecc_{\rm max}$ in \cref{fig:med-max-ecc}. For comparison, a pessimistic evaluation of \cref{eqn:max-g-ecc} using the maximum frequency and chirp mass in each configuration results in a sensitivity to binaries with maximum eccentricities of $[0.33,0.72]$, with a median of $0.46$. On the other hand, an optimistic evaluation using the minimum frequency and chirp mass probed in each configuration would permit us to detect binaries with eccentricities between $[0.86,0.97]$, with a median of $0.91$.

\begin{figure}[ht!]
    \centering
    \includegraphics[width =\linewidth]{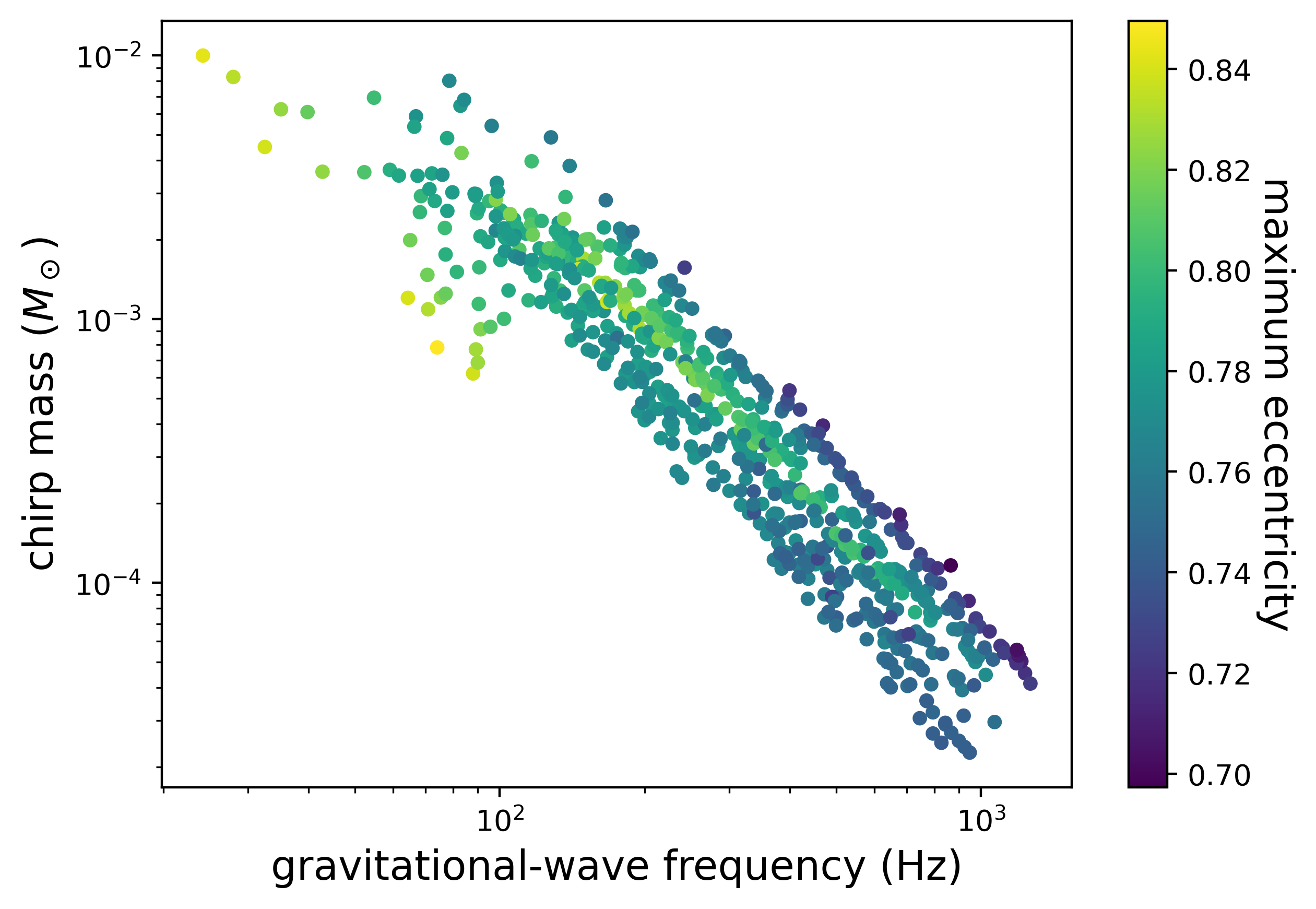}
    \caption{\textbf{The median maximum eccentricity that each configuration in our search could probe}. We show that, for asymmetric mass-ratio systems, we can be sensitive to eccentricities as high as 0.84, depending on the \gwh frequency and chirp mass of the system.}
    \label{fig:med-max-ecc}
\end{figure}

\subsection{Conditions on the PBH mass distribution}

Calculating \pbh merger rates is highly model-dependent. In particular, the choice of the \pbh mass distribution can significantly affect the predicted suppression of merger rates. This suppression mainly arises from the disruption of binaries, either through early interactions with intruding \pbhs or via late encounters with other \pbhs within clusters formed by Poisson fluctuations. Some sufficient conditions to avoid such a suppression have been identified, and in this appendix we provide a more detailed explanation for these conditions, based on the current prescriptions for the merger rates. 

\subsubsection{Early perturbers}

The first two conditions ensure that most binaries are not gravitationally bound to their neighbors when forming. Assuming $f_{\rm PBH} \gg \sigma_{\rm M}$ (where $\sigma_{\rm M}^2$ is the variance of matter density perturbations at the time the binary forms), the suppression factor associated with the early disruption of binaries by a nearby intruder can be expressed as \cite{Raidal:2018bbj, Hutsi:2020sol,Raidal:2024bmm}: 
\begin{equation}
    f_{\rm sup}^{(\rm dist)} \approx 1.42 \left(\frac{\langle m ^2 \rangle / \langle m\rangle ^2}{\bar{N}(y) + 0.4}    \right)^{- 21/74} \exp({-\bar{N})},
    \label{supp_raidal}
\end{equation}
where $\bar{N}$ is the expected number of \pbhs within the binary's sphere of influence at scale factor $a$, given by 
\begin{equation}
    \bar{N} = \frac{M}{\langle m \rangle}\frac{f_{\rm PBH}}{f_{\rm PBH} + a_{\rm eq}/a},
\end{equation}
with $M$ denoting the total mass of the binary, $\langle m \rangle = \rho_{\rm PBH} / n_{\rm PBH}$  the average \pbh mass, and $a_{\rm eq}$ the scale factor at matter-radiation equality. Since the number density of \pbhs, $n_{\rm PBH}$, scales inversely with mass, lighter \pbhs can significantly outnumber the \pbhs from the peak, even with a subdominant contribution to the total \pbh energy density, increasing $\bar{N}$ and driving the suppression factor towards zero. Consequently, only mass functions with a suppressed low-mass tail can survive such heavy disruption. In these cases, for symmetric binaries, one typically finds $\bar{N}\approx 2$ yielding a suppression factor of $f_{\rm sup} \approx 0.2$.  For asymmetric binaries, $M \approx m_1$ such that $\bar N \approx 1$, which yield a suppression factor of $f_{\rm sup} \approx 0.5$.  These are the two  cases we consider to produce  Fig.~\ref{fig:money}.


\subsubsection{Late perturbers}

One additional condition that must be fulfilled in all scenarios is to avoid the rate suppression due to the interactions of the binaries with \pbhs inside clusters seeded by Poisson fluctuations.  The standard prescription, validated by numerical simulations for peaked distributions, leads to a rate suppression factor today approximately given by 
\begin{equation}
f_{\rm sup}^{\rm(cl)} \approx 0.01 f_{\rm PBH}^{-0.65}
\end{equation}
when a significant \dmh fraction is made of \pbhs, i.e. an additional suppression by up to two orders of magnitude.  However, we can show that such a suppression can be avoided for low-mass \pbhs.  From Ref.~\cite{Raidal:2024bmm}, binaries are perturbed by encounters with \pbhs of mass $m_3$ inside clusters when the impact parameter is 
\begin{equation}
b \lesssim \frac{\left[m_3 (m_1+m_2+m_3)\right]^{1/3} r_a^{1/2} }{j_\tau^{1/3} (m_1 + m_2)^{1/6} v_{\rm rel}}
\end{equation}
where $r_a$ is the separation and $j_\tau$ the angular momentum of the binary, expected to merge today, for which estimations are provided in Ref.~\cite{Raidal:2024bmm},
\begin{equation}
j_\tau = 1.7 \times 10^{-2} \left(\frac{m_1+m_2}{\msun}\right)^{5/37} \frac{(4 m_1 m_2)^{3/37}   }{(m_1+m_2)^{6/37}} f_{\rm PBH}^{16/37}
\end{equation}
and 
\begin{equation}
r_a = 2.13 \times 10^9 {\rm m} \times \left[ \frac{m_1 m_2 (m_1+m_2)}{\msun^3} \right]^{1/4}~,
\end{equation}
and where $v_{\rm rel}$ is the relative velocity, of same order than the virial velocity of the cluster.  The corresponding interaction cross-section is then given by $\sigma = \pi b^2$ and one can then estimate, for each \pbh mass $m_3$ with an abundance characterized by $f(\ln m_3)$, the typical time it takes to perturb the binary, $t_{\rm pert} = 1/[n(m_3) \sigma v_{\rm rel}]$, where $n(m_3)$ is the number density of perturbers of mass $m_3$ in the cluster.  For a wide mass distribution, one has to integrate $n(m_3) \sigma v_{\rm rel}$ over the full distribution to get the total perturbation rate.  For simplicity, we only show here that any value of $m_3$ gives rise to $t_{\rm pert} \gg t_0$, the age of the Universe, with some conditions on the \pbh clusters.  But in general, $t_{\rm pert}$ depend on the typical clustering scale.  From Ref~\cite{Raidal:2024bmm}, this time was found to be typically much smaller than the age of the Universe, leading to automatic binary perturbations in clusters.  Nevertheless, the clustering process is complex and is also impacted by the dynamical heating of small clusters and their dilution in larger clusters.  As explained in~\cite{Carr:2023tpt}, if there exist a mass range $m$ in the distribution for which one has $f(m) m / \msun \sim \mathcal O(1)$, then the smallest clusters will have a radius of order 20 parsecs and a mass around $10^6 \,\msun$ that could be associated to ultra-faint dwarf galaxies, a larger value of that combination being excluded by the observations of these ultra-faint dwarfs.  These clusters would have virial velocities of order of a few km/s.  We have calculated the interaction time needed to perturb binaries with component masses $m_1$ and $m_2$ as
\begin{align}
t_{\rm pert} & \approx 2.7 \times 10^{10} {\rm yr} \times f_{\rm PBH}^{-79/111} f(\ln m_3)^{-1} \nonumber \\ 
&\times \left(\frac{m_1 m_2}{\msun^2}\right)^{-\frac{29}{148}} \left(\frac{m_3}{\msun} \right)^{\frac 1 3} \left( \frac{m_1+m_2}{\msun} \right)^{\frac{29}{144}} \nonumber \\
& \times \left( \frac{m_1+m_2+m_3}{\msun} \right)^{-\frac 2 3}  \left( \frac{M_{\rm cl}}{\msun} \right)^{-\frac 1 2} \left( \frac{r_{\rm cl}}{\rm pc} \right)^{5/2}~,
\end{align}
where $M_{\rm cl}$ and $r_{\rm cl}$ are the cluster mass and radius, respectively.  Taking $f(\ln m_3) =1$, for the typical above-mentioned clusters and the relevant masses and values of $f_{\rm PBH}$ for this work, this time is found to exceed the age of the universe, lower abundances leading obviously to even larger interaction times.  But let us note again the importance of the large cluster mass compared to what was considered in~\cite{Hutsi:2020sol,Raidal:2024bmm}. 
A heavy-mass tail in the distribution, or of a high peak at the solar-mass scale with a significant $f_{\rm PBH}$, are therefore required to prevent merger rates to be suppressed due to \pbh clusters.

{Finally, let us notice that we have checked that the other \pbh binary formation channels, reviewed in~\cite{Raidal:2024bmm}, namely the 3-body early binary formation channel and the late 2-body and 3-body channels in \pbh clusters, generically lead to lower merger rates than for early binaries, for significant values of $f_{\rm PBH}$, both for equal-mass subsolar mergers and for asymmetric binaries.  These have therefore not been considered in our analysis.  }

\section*{Acknowledgments}

This material is based upon work supported by NSF's LIGO Laboratory, which is a
major facility fully funded by the National Science Foundation.
The authors also gratefully acknowledge the support of
the Science and Technology Facilities Council (STFC) of the
United Kingdom, the Max-Planck-Society (MPS), and the State of
Niedersachsen/Germany for support of the construction of Advanced LIGO 
and construction and operation of the GEO\,600 detector. 
Additional support for Advanced LIGO was provided by the Australian Research Council.
The authors gratefully acknowledge the Italian Istituto Nazionale di Fisica Nucleare (INFN),  
the French Centre National de la Recherche Scientifique (CNRS) and
the Netherlands Organization for Scientific Research (NWO)
for the construction and operation of the Virgo detector
and the creation and support  of the EGO consortium. 
The authors also gratefully acknowledge research support from these agencies as well as by 
the Council of Scientific and Industrial Research of India, 
the Department of Science and Technology, India,
the Science \& Engineering Research Board (SERB), India,
the Ministry of Human Resource Development, India,
the Spanish Agencia Estatal de Investigaci\'on (AEI),
the Spanish Ministerio de Ciencia, Innovaci\'on y Universidades,
the European Union NextGenerationEU/PRTR (PRTR-C17.I1),
the ICSC - CentroNazionale di Ricerca in High Performance Computing, Big Data
and Quantum Computing, funded by the European Union NextGenerationEU,
the Comunitat Auton\`oma de les Illes Balears through the Conselleria d'Educaci\'o i Universitats,
the Conselleria d'Innovaci\'o, Universitats, Ci\`encia i Societat Digital de la Generalitat Valenciana and
the CERCA Programme Generalitat de Catalunya, Spain,
the Polish National Agency for Academic Exchange,
the National Science Centre of Poland and the European Union - European Regional
Development Fund;
the Foundation for Polish Science (FNP),
the Polish Ministry of Science and Higher Education,
the Swiss National Science Foundation (SNSF),
the Russian Science Foundation,
the European Commission,
the European Social Funds (ESF),
the European Regional Development Funds (ERDF),
the Royal Society, 
the Scottish Funding Council, 
the Scottish Universities Physics Alliance, 
the Hungarian Scientific Research Fund (OTKA),
the French Lyon Institute of Origins (LIO),
the Belgian Fonds de la Recherche Scientifique (FRS-FNRS), 
Actions de Recherche Concert\'ees (ARC) and
Fonds Wetenschappelijk Onderzoek - Vlaanderen (FWO), Belgium,
the Paris \^{I}le-de-France Region, 
the National Research, Development and Innovation Office of Hungary (NKFIH), 
the National Research Foundation of Korea,
the Natural Sciences and Engineering Research Council of Canada (NSERC),
the Canadian Foundation for Innovation (CFI),
the Brazilian Ministry of Science, Technology, and Innovations,
the International Center for Theoretical Physics South American Institute for Fundamental Research (ICTP-SAIFR), 
the Research Grants Council of Hong Kong,
the National Natural Science Foundation of China (NSFC),
the Israel Science Foundation (ISF),
the US-Israel Binational Science Fund (BSF),
the Leverhulme Trust, 
the Research Corporation,
the National Science and Technology Council (NSTC), Taiwan,
the United States Department of Energy,
and
the Kavli Foundation.
The authors gratefully acknowledge the support of the NSF, STFC, INFN and CNRS for provision of computational resources.

This work was supported by MEXT,
the JSPS Leading-edge Research Infrastructure Program,
JSPS Grant-in-Aid for Specially Promoted Research 26000005,
JSPS Grant-in-Aid for Scientific Research on Innovative Areas 2402: 24103006,
24103005, and 2905: JP17H06358, JP17H06361 and JP17H06364,
JSPS Core-to-Core Program A.\ Advanced Research Networks,
JSPS Grants-in-Aid for Scientific Research (S) 17H06133 and 20H05639,
JSPS Grant-in-Aid for Transformative Research Areas (A) 20A203: JP20H05854,
the joint research program of the Institute for Cosmic Ray Research,
University of Tokyo,
the National Research Foundation (NRF),
the Computing Infrastructure Project of the Global Science experimental Data hub
Center (GSDC) at KISTI,
the Korea Astronomy and Space Science Institute (KASI),
the Ministry of Science and ICT (MSIT) in Korea,
Academia Sinica (AS),
the AS Grid Center (ASGC) and the National Science and Technology Council (NSTC)
in Taiwan under grants including the Science Vanguard Research Program,
the Advanced Technology Center (ATC) of NAOJ,
and the Mechanical Engineering Center of KEK.

Additional acknowledgements for support of individual authors may be found in the following document: \\
\texttt{https://dcc.ligo.org/LIGO-M2300033/public}.
For the purpose of open access, the authors have applied a Creative Commons Attribution (CC BY)
license to any Author Accepted Manuscript version arising.
We request that citations to this article use 'A. G. Abac {\it et al.} (LIGO-Virgo-KAGRA Collaboration), ...' or similar phrasing, depending on journal convention.




\bibliographystyle{apsrev4-1}
\bibliography{references} 

\iftoggle{endauthorlist}{
  \let\author\myauthor
  \let\affiliation\myaffiliation
  \let\maketitle\mymaketitle
  \title{The LIGO Scientific Collaboration, Virgo Collaboration, and KAGRA Collaboration  }
  \pacs{}
%

\author{A.~G.~Abac\,\orcidlink{0000-0003-4786-2698}}
\affiliation{Max Planck Institute for Gravitational Physics (Albert Einstein Institute), D-14476 Potsdam, Germany}
\author{I.~Abouelfettouh}
\affiliation{LIGO Hanford Observatory, Richland, WA 99352, USA}
\author{F.~Acernese}
\affiliation{Dipartimento di Farmacia, Universit\`a di Salerno, I-84084 Fisciano, Salerno, Italy}
\affiliation{INFN, Sezione di Napoli, I-80126 Napoli, Italy}
\author{K.~Ackley\,\orcidlink{0000-0002-8648-0767}}
\affiliation{University of Warwick, Coventry CV4 7AL, United Kingdom}
\author{C.~Adamcewicz\,\orcidlink{0000-0001-5525-6255}}
\affiliation{OzGrav, School of Physics \& Astronomy, Monash University, Clayton 3800, Victoria, Australia}
\author{S.~Adhicary\,\orcidlink{0009-0004-2101-5428}}
\affiliation{The Pennsylvania State University, University Park, PA 16802, USA}
\author{D.~Adhikari}
\affiliation{Max Planck Institute for Gravitational Physics (Albert Einstein Institute), D-30167 Hannover, Germany}
\affiliation{Leibniz Universit\"{a}t Hannover, D-30167 Hannover, Germany}
\author{N.~Adhikari\,\orcidlink{0000-0002-4559-8427}}
\affiliation{University of Wisconsin-Milwaukee, Milwaukee, WI 53201, USA}
\author{R.~X.~Adhikari\,\orcidlink{0000-0002-5731-5076}}
\affiliation{LIGO Laboratory, California Institute of Technology, Pasadena, CA 91125, USA}
\author{V.~K.~Adkins}
\affiliation{Louisiana State University, Baton Rouge, LA 70803, USA}
% ...



  \newpage
  \maketitle
}

\end{document}